\documentclass[prl,twocolumn,aps,superscriptaddress]{revtex4}
\usepackage{amsmath}
\usepackage{graphicx}
\usepackage{tabularx}
\usepackage{amssymb}

\begin{document}

\title{Lighting the Landscape: Molecular Events Under Dynamic Stark Shifts} 

\author{Bo Y. Chang}
\affiliation{School of Chemistry (BK21), Seoul National University, Seoul 151-747, Republic of Korea}

\author{Ignacio R. Sola}
\affiliation{Departamento de Qu\'imica F\'isica, Universidad Complutense, 28040 Madrid, Spain}
\email{isola@quim.ucm.es}

\author{Seokmin Shin}
\affiliation{School of Chemistry (BK21), Seoul National University, Seoul 151-747, Republic of Korea}
\email{sshin@snu.ac.kr}

\begin{abstract}
A new perspective on how to manipulate molecules by means of very
strong laser pulses is emerging with insights from the so-called
light-induced potentials, which are the adiabatic
potential energy surfaces of molecules severely distorted by
the effect of the strong field.
Different effects appear depending on how the laser frequency
is tuned, to a certain electronic transition, creating light-induced
avoided crossings, or very off-resonant, generating Stark shifts.
In the former case it is possible to induce dramatic
changes in the geometry and redistribution of charges in the molecule 
while the lasers
are acting and to fully control photodissociation reactions as well as 
other photochemical processes.
Several theoretical proposals taken from the work of the authors are 
reviewed and analyzed showing the unique features that the 
strong-laser chemistry opens to control the transient properties and 
the dynamics of molecules.
\end{abstract}

\maketitle



\section{Introduction} 


In the last few years, many techniques were developed to manipulate and
control molecular processes by means of ultrashort laser pulses\cite{OCT1,OCT2,OCT3,OCT4,OCT5}.
Initially, ultrashort pulses were used because they provided the means 
to act on the time-scale of fast molecular events\cite{Zewail}. 
At a later stage, it was the broad pulse spectrum 
of the pulses that by phase modulation and learning algorithms, opened 
great opportunities to control the dynamics\cite{shaping}.
More recently, the non-resonant strong-field interaction of the pulse provided
a means to alter the potential energy surface of the molecule via dynamic
Stark effect, ''catalyzing'' many photochemical 
processes\cite{SFC1,SFC2,SFC3,SFC4}.
While all the previous roles remain useful for different purposes,
one can arguably
follow this sequence of events as one in which the laser is promoted from the role of a non-specific exciter (and probe of the chemical process), to a reactant and then to a catalyst, using the chemical terminology.
In this review we will outline some of our contributions in the control
of molecules in the strong-field regime from a theoretical perspective.
For a broader perspective the interested reader is referred to the following
work and references therein\cite{SFCR1,SFCR2,SFCR3}. 

The application of strong pulses to atoms has a long history\cite{atoms1,atoms2,atoms3,atoms4,atoms5,atoms6,atoms7,atoms8,atoms9,atoms10,atoms11}.
Here we are interested in non-ionizing effects, which require the use
of moderately-strong pulses, typically within tens of TW/cm$^2$.
Thomas George and Andre Bandrauk developed a very useful ''chemical'' 
picture of light-induced events\cite{George,Bandrauk}. 
In this picture, the slow effects of the field
on the nuclei (averaged over the radiation cycles) are incorporated in
the ''dressed'' (energy shifted) molecular potentials. Then, in the adiabatic
representation, the dynamics of photodissociation or multiphoton processes 
are recast in terms of predissociation, avoided crossings, or other 
topological features.
The adiabatic potentials that incorporate these laser-molecule coupling
effects
are called light-induced
potentials\cite{LIPs} (LIPs), while the avoided crossings in the LIPs are called
light-induced avoided crossings\cite{LIAC1,LIAC2} (LIACs). 
In polyatomic molecules, 
or taking into account the vectorial nature of the coupling, the LIACs
can be seen as light-induced conical intersections\cite{LICI1,LICI2,LICI3,LICI4,LICI5,LICI6,LICI7,LICI8,LICI9} (LICIs).
Considerable theoretical effort has been put recently at characterizing the
LICIs.

One of the first and most obvious applications of strong pulses is to 
enhance the yield of electronic absorption. However, because Raman 
transitions can compete with 
absorption and the Stark effect
can decouple the electronic states, strong and ultrashort 
transformed-limited pulses do not lead to efficient population transfer\cite{Par0,Par1,Par2,Par3}.
One needs to resort to adiabatic rapid passage (ARP) by means of
chirped pulses, or to schemes that use more than one pulse\cite{ARP1,ARP2,ARP3,ARP4,STIRAP}.
Simple theoretical models, such as the Landau-Zener model, were often
used to explain the high yields of electronic excitation in ARP\cite{GarSuorev}.
On the other hand, several schemes for population transfer with sequences
of strong pulses could be designed as strong-field analogs in electronic
potentials of adiabatic passage between quantum levels\cite{Vitanov}.
Adiabatic passage by light-induced potentials\cite{APLIP1,APLIP2,APLIP3,APLIP4,APLIP5,APLIP6,APLIP6b,APLIP7} (APLIP),
chirped adiabatic rapid passage\cite{CARP1,CARP2,CARP3} (CARP), 
selective population of dressed states\cite{SPODS1,SPODS2,SPODS3,SPODS4} (SPODS), 
rapid vibrational inversion via time-gating\cite{LIAC1}
or Raman chirped adiabatic passage\cite{RCAP1,RCAP2,RCAP3,RCAP4,RCAP5,RCAP6} (RCAP) serve as examples.

The LIPs are not only useful as a convenient explanatory device to understand 
the remarkable features of population transfer under strong pulses 
(e.g., its robutness); they truly are potential energy surfaces that determine, 
for instance, the geometrical features of the molecule 
and their related properties.
Particularly interesting are LIPs formed by coupling a bound and a dissociative
molecular potential. The effect of a strong field is to mix their electronic
character.
The first evidence of these properties was found out after observing bond softening
in the ground electronic state\cite{soft1,soft2,soft3,soft4,soft5,soft6}
 and bond hardening in a dissociative potential\cite{hard1,hard2,hard3,hard4,hard5}.
Varying the intensity and frequency of the dressing field one can
efficiently control the geometry of the ''previously-dissociating'' potential. 

However, in order to change
the molecular properties, it is necessary to adiabatically prepare the system 
on this potential. One can then externally control the bond distance of
a diatomic molecule over a very large range of values, as in the laser adiabatic
manipulation of the bond (LAMB) scheme.
Several two-pulse schemes and single pulses with modulated frequencies were 
proposed for this purpose\cite{LAMB1,LAMB2,LAMB3,LAMB4,LAMBc1,LAMBc2,LAMBc3,LAMBc4}.
If the preparation is not fully adiabatic, one can still control the
transition to create oscillating nuclear wave packets of different amplitudes
in the LIP, that is, to create molecular analogs of ''classical-like''
coherent vibrations\cite{LAMBc2,LAMBc4}. 
It is also possible to correct the non-adiabaticity 
by ”absorbing” the excess of vibrational energy as a zero-energy of 
a modified LIP, via time-asymmetric pulses\cite{LAMBcor}.
In addition to controlling the molecular geometry, other LIPs can be prepared
to change the width of the nuclear wave packet, achieving molecular
squeezing either adiabatically or dynamically\cite{squeez1,sqz1,sqz2,squeez2,squeez3}.
The adiabatic methods in principle can be used to create {\it artificial}
bond lengths and vibrations with parameters that are fully externally
controlled by the laser. However, these properties are {\it transient}. They
only exist as long as the laser is acting and no strong measurement is
performed on the system.

In correspondence to the geometrical changes induced by the LIP,
there are changes in the electronic properties associated to
the electronic superposition state\cite{eLIP1,eLIP2,eLIP3,eLIP4}. 
These properties, for example
the permanent or transition dipole, reflect the underlying changes
in the redistribution of charges. It was recently shown that
some superpositions manifested a clear classical picture of
an electron oscillating between the protons, whereas in a dressed
electronic state the electron was moving along with the proton\cite{eLIP3}.

The electronic character of the LIP plays important roles in other
processes as well. For instance, in the control of the spin state
of the molecule, we have shown how electric pulses or electromagnetic
fields can be used to avoid a singlet-triplet transition, by 
suitably modifying the singlet and triplet LIPs such that there 
is no crossing between them\cite{spin1,spin2,spin3}.
However, under usual circumstances very strong fields are needed,
such that the rate of ionization at the required laser intensity is 
faster than the rate of inter-system conversion\cite{spinion1,spinion2}. 
Under certain conditions,
the schemes can only operate when the spin-orbit coupling is weak,
reverting to a few-level problem\cite{spinweak,spinion2}.

Other intramolecular or non-adiabatic couplings can be controlled in
a similar manner. For instance, one can generate LIACs that prevent the
initial nuclear wave packet to reach a certain conical intersection
in order to avoid energy deactivation\cite{LIAC2}. However, most works have 
been done to create the LIAC or LICI in order to control the output of a 
photochemical process. 
In particular, there have been several theoretical proposals to 
control the yield of a photodissociation reaction\cite{diss3}, as well as
the branching ratio over possible fragmentation channels\cite{diss0,diss2,diss3,LICI5} and
the kinetic energy distribution of the fragments\cite{diss4}.
Although most proposals remain experimentally untested, owing to the difficulty
of finding good molecular systems where the strong field interaction is
strong enough to generate LIPs, yet not too strong that the ionization
is predominant, recent experiments have shown that indeed such control is 
possible\cite{SFC1,SFC2,SFC3}.

In this work we will review some of our findings.
In Sec. 2 we will provide a simplified analysis of the geometrical and
dynamical features of LIPs, outlining the role of the Stark-shift
and of laser-induced potential energy shaping in several control scenarios.
In Sec. 3 we present several works of our group for the control
of the bond length in diatomic molecules, by using LIPs that imply 
contributions of bound and dissociative electronic states.
In Sec. 4 we analyze the electronic character of the LIPs and the
interesting views that it provides to control electronic properties such
as the dipole moment. In Sec. 5 we show how one can control different
observables of photodissociation reactions, such as the photodissociation
spectrum, the branching ratios and the kinetic energy distribution of
the fragments, by using strong nonresonant fields that couple two
dissociating electronic states. Finally, Sec. 6 provides some of the relevant
Conclusions.

\section{Light-Induced Potentials: Geometrical and Dynamical Features}

The goal of this section is to set the stage where all the subsequent
control schemes reviewed in this article operate, clarifying 
the relation between certain topological features of the LIPs and the 
quantum processes that they convey.
To simplify the analysis we start by considering diatomic molecules 
oriented with respect to a single external field $E(t)$.
We use the rotating wave approximation (RWA), such that 
$E(t) \approx {\epsilon}(t) e^{\pm i\varphi(t)}/2$,
where ${\epsilon}(t)$ is a slowly varying envelope function, compared to 
the rate of change of the dynamical phase $\varphi(t)$. 
The negative sign is used to describe
absorption, while the positive sign is used for the stimulated emission.
In general, for chirped pulses, $\varphi(t) = \int \omega(t')dt'$
where $\omega(t)$ is the time-varying frequency.
If we assume that only two electronic states participate in the dynamics,
the following very general effective Hamiltonian can be used to describe
the evolution of the nuclear wave functions in each electronic state,
\begin{widetext}
\begin{equation}
{\sf \bf{H}} = \left( \begin{array}{cc} {\sf T} & {\sf K} \\ {\sf K} & {\sf T} 
\end{array} \right) + \left( \begin{array}{cc} 
V_1(R) - \frac{1}{4} \alpha_{11}(R) {\epsilon}^2(t) &
-\frac{1}{2} \mu_{12}(R) {\epsilon}(t) \\ 
-\frac{1}{2} \mu_{12}(R) {\epsilon}(t) & 
V_2(R) - \hbar\omega(t) - \frac{1}{4} \alpha_{22}(R) {\epsilon}^2(t) 
\end{array} \right)  \label{H2Ppol}
\end{equation}
\end{widetext}
where ${\sf T}$ is the kinetic energy, ${\sf K}$ takes into account 
non-adiabatic couplings and $V_j(R)$ are the electronic potential energy 
curves. 
The dynamical phases in the coupling (off-diagonal term in the Hamiltonian)
are moved by a unitary transformation to the energies, showing photon-shifted
potentials.
In Eq.(1) we have assumed that the field may be resonant or
quasi-resonant between the two electronic states, coupled via the dipole
moment $\mu_{12}$, and nonresonant
with respect to the remaining states of the molecule. 
The effect of the remaining states on the two selected states
is described in terms of the quasi-polarizabilities 
($\alpha_{11}$ and $\alpha_{22}$)
up to the next leading order in the field, ${\epsilon}^2$. They account for
the Stark shifts. In principle, two very different fields (with very
different frequencies) could be responsible for the electronic coupling and
the Stark-shifts. In more general scenarios, the quasi-resonant electronic
transition could require multiphoton absorption instead of the single
photon absorption used in Eq.(1).

We consider now two different regimes depending on whether the effect
of the laser on
the potentials renders a ''soft'' or ''hard'' shaping.
The first one is characterized by the lack of a resonant or quasi-resonant
excitation so that the off-diagonal terms are negligible. Then the
initially populated LIP can be written as
$$ V_1^{a}(R,{\epsilon}) \approx V_1(R) - \frac{1}{4} \alpha_{11}(R) {\epsilon}^2
(t)$$
in which $\alpha$ is the dynamic polarizability, taking into account
the effect of all the remaining states. In some cases the
polarizability is dominated by a single electronic state, closer in energy
to $V_1(R) + \hbar\omega$. In other cases, the frequency is much smaller
({\it e.g.} an infrared laser or an electric pulse) and the static 
polarizability can be used instead. Unless $\alpha_{11}(R)$ changes
drastically around the equilibrium geometry of $V_1(R)$ (or the regions
where the probability of finding the nuclear wave function are larger),
the topological changes in $V_1(R)$ induced by the field will be
small, hence the ''soft'' character of the shaping. The control is
mainly exerted by ${\epsilon}(t)$, inducing energy variations (Stark shifts)
of the potential. It is often the case that the ground LIP is very
similar to the ground molecular potential. Most interesting effects
occur in excited LIPs. One first needs 
to move the population to an excited potential such that the events
happen in $V_2^{a}$.

On the other hand, when the interaction is quasi-resonant, as a first
approximation one can typically neglect the polarizability and
concentrate on the two crossing potentials.
The LIPs are obtained by diagonalizing the potential energy operator,
including the field coupling. They are the instantaneous eingenstates of
the electronic Hamiltonian.
Applying the rotation matrix
$$\left( \begin{array}{cc} \cos\theta(R;\epsilon) & \sin\theta(R;\epsilon) \\
-\sin\theta(R;\epsilon) & \cos\theta(R;\epsilon) \end{array} \right)$$
where $\theta(R;\epsilon)$ is the rotation or mixing angle that diagonalizes
the matrix,
we obtain
\begin{equation}
{\sf \bf{H}}^\mathrm{DS} =
\left( \begin{array}{cc} {\sf T} & {\sf K}' \\ {\sf K}' & {\sf T} \end{array} 
\right) 
+ \left( \begin{array}{cc} 
V^{a}_1(R;{\epsilon}) & i\dot{\theta}\cos2\theta \\ -i\dot{\theta}\cos2\theta 
& V^{a}_2(R;{\epsilon}) \end{array} \right)  \label{HLIP}
\end{equation}
The off-diagonal terms in the kinetic operator, ${\sf K}'$  are often referred 
to as {\em spatial} non-adiabatic terms, while those in the potential operator
are {\em dynamical} non-adiabatic terms. 
They depend on the time-derivative in the mixing angle, $\dot{\theta}$, 
which reflects the time-variation of the field, $\dot{\epsilon}(t)$.
When the pulses are strong enough and their time evolution is slow enough
(in comparison with the motion of the nuclear wave functions) the 
off-diagonal terms can be neglected.
Then, if at initial time (when $\epsilon(0) = 0$) the initial potential
correlates with a single LIP, $V_1^{a}(R;\epsilon(0))$, all the
dynamics will occur in this LIP and the final electronic state as well
as all the properties of the system during all times, will solely depend
on $V_1^{a}(R;\epsilon(t))$.
In order to characterize the LIP we need to know the structure of the
strongly coupled electronic potentials, $V_1$ and $V_2$.
It is most important to localize the LIAC, $R_c$, defined by the condition
\begin{equation}
\Delta(R_c,t) = V_2(R_c) - V_1(R_c) - \hbar\omega(t)  = 0 \ .
\end{equation}
The populated LIP can be expressed as a function of the original molecular potentials, as
\begin{equation}
V^{a}_1(R,{\epsilon}) = \cos\theta(R,{\epsilon})\, V_1(R) 
+ \sin\theta(R,{\epsilon}) \,V_2(R)  \label{VLIP}
\end{equation}
where the mixing angle $\theta$ 
changes from $0$ to $\pi/2$ at both sides of the avoided crossing, $R_c$.
The first important effect
that such an avoided crossing has in the LIPs is to completely deform
the molecular potential energy curves and thus to change the structure of
the molecule.
For $V^{a}_1(R)$ the LIP looks like $V_1(R<R_c)$ before the crossing and
like $V_2(R>R_c)$ after it.
Through the LIAC, the nuclear wave packet can transfer part of
the population. It operates in analogous way to a molecular
(beyond Born-Oppenheimer-like) internal conversion, induced by ${\sf K}$. 
In the adiabatic limit, which requires a large energy gap between the
LIPs in the LIAC and slow changes in the pulse envelope ${\epsilon}(t)$,
the population in the initial electronic
state is given by $\cos^2\theta(R,{\epsilon})$, while the population
in the other coupled electronic state is given by $\sin^2\theta(R,{\epsilon})$.
Therefore, the motion of a nuclear wave packet across $R_c$ in 
$V_1^{a}(R,{\epsilon})$ represents full population transfer from $V_1$ to $V_2$.

We will now briefly mention some features of population transfer analyzed 
from the perspective of LIPs.
As noticed,
one of the most important steps in the design of laser control schemes is 
to localize the LIAC of the LIP, as this topological point
is an indication of possible population inversion.
In order to fully transfer the population from $V_1$ to $V_2$
one needs to modulate $\theta$ via the control field ${\epsilon}(t)$.
However, depending on the structure of the LIP and the initial kinetic energy,
the nuclear wave function will be able or not to cross the region of 
the potential that correlates with $V_2$.
In the most simple cases, as {\it e.g.} in population transfer from a bound
to a dissociating electronic state, a chirped
pulse where the pulse frequency $\omega(t)$ sweeps through the Franck-Condon
region is often enough to allow the mixing angle $\theta(R,{\epsilon})$ to 
change from $0$ to $\pi/2$ for all values of $R$ where the wave packet is
located. 
Therefore, most LAMB schemes use chirped pulses. 
In other cases, one needs to find a more difficult adiabatic path that
connects $V_1$ to $V_2$ via the LIP, requiring a more elaborate
trajectory of $\theta(R,{\epsilon})$. Typically, when the equilibrium
geometries of $V_1$ and $V_2$ are very separated and the energy gap between
the LIPs at the LIAC is large,
one needs to find additional electronic states that allow to modulate
the LIP from $V_1$ to $V_2$ adiabatically. 
The APLIP scheme using time-delayed pulse sequences, or the CARP scheme,
using chirped pulses, control the population inversion to a higher excited state
by means of two-photon absorption.
This requires the use of
more than one control pulse, therefore adding challenges to the
experimental implementation of the scheme.

If the population transfer is fully adiabatic, there is no internal
barrier in the adiabatic pathway at the bottom of the LIP connecting 
the initial equilibrium geometry corresponding to $V_1$ and the final
equilibrium geometry corresponding to $V_2$. Under these circumstances
the transfer preserves the form of the nuclear wave function. In particular,
the dynamics conserves the vibrational quanta.
In APLIP, this is possible by using two control pulses, one called
the pump pulse ${\epsilon}_p(t)$, the second one called the Stokes pulse
${\epsilon}_S(t)$. For instance, consider that we want to invert the
population in Na$_2$ from the ground X$^1\Sigma^+_g$ state ($V_1$) to a second
excited 2$^1\Pi_g$ state ($V_2$), using a resonant two-photon transition through
the intermediate A$^1\Sigma_u^+$ state ($V_b$), whose equilibrium geometry lies
in between that of the initial and the final state 
(although this is not an essential requirement for the intermediate state,
it typically reduces the pulse intensities needed for the APLIP scheme).
Garraway and Suominen\cite{APLIP1} showed that a counter-intuitive pulse sequence with 
${\epsilon}_S(t)$ preceding ${\epsilon}_p(t)$ could lead to full population 
inversion without populating the intermediate state at all.
This is possible because such pulse sequence prepares a LIP called $V^{a}_d$,
$V^{a}_d = \cos\theta V_1 - \sin\theta V_2$,
that never correlates with the intermediate potential $V_b$.
Also interestingly, in this LIP the Stark effects induced by both control
pulses are minimal regardless of the pulse strength.
With different properties, other APLIP pulse sequences allow full adiabatic
passage\cite{APLIP4,APLIP5}.

In the following subsections we will analyze several examples of theoretically
proposed schemes of control exerted via Stark effect (that is, when the
coupling is far off-resonant and at least as a first approximation one
can use the polarization), and control exerted by LIP shaping
(when the coupling is resonant or quasi-resonant and one has
to take into consideration substantial population transfer).
In the latter, the geometrical factors are obvious, as the population
transfer is typically encoded in the reshaping of the LIP, but
the dynamics play also a very important role on the creation, passage
or destruction of the LIPs.

\section{Control of Molecular Geometry by Shaping Bound with Dissociative
Potentials}

In this section we will review several schemes based on LIPs
that were designed to control the bond length of diatomic molecules.
In general, a molecule has different equilibrium geometries in each 
electronic state. Typically, the bonds are more relaxed in the 
excited states. Therefore, by electronic absorption with strong 
ultrashort pulses, it is possible to transfer all the initial wave 
function to the excited state, creating a wave packet of vibrational
eigenstates after the pulse is switched off. This vibrational wave packet
oscillates around the excited-state equilibrium geometry of $V_{2}$ for a 
few periods so that the bond length of the molecule will be well defined
until dispersion occurs. As the dynamics is driven by the molecular potential,
there is little control over the period or amplitude of the motion.
 
Alternatively, the electronic absorption may proceed adiabatically, as
in APLIP. Then the nuclear wave function will be slowly transferred 
to the equilibrium geometry of the second excited electronic state.
During this process, at every intermediate time the bond length is
well defined, as the wave function sits in the bottom of a single LIP.
The idea behind the LAMB scheme is precisely to stop or freeze the
transport process at the desired intermediate bond length. As long as the pulses
remain constant, so will the bond length.  
However, the potential range for the control is much larger when the final 
electronic state is dissociative, since then its equilibrium 
geometry is at $R\rightarrow \infty$. Hence, in principle, one 
can prepare the molecule at any bond length larger than the initial one.
However, the adiabaticity of the process deteriorates for very large
bond distances, and the dissociation probability becomes non-negligible.

The first proposed implementation of LAMB assumed that the initial potential 
$V_1$ and the intermediate potential $V_3$, 
where both bound potentials, coupled via a pump
pulse $\epsilon_1(t)$, while the target electronic state $V_2$ was dissociative, 
coupled to the intermediate one via $\epsilon_2(t)$. 
Following one possible APLIP sequence but shaping $\epsilon_2(t)$ such that 
instead of switching it off, the pulse remains at a plateau amplitude
$\epsilon_0$ for a certain time,
it was possible to elongate the bond\cite{LAMB1}.

\begin{figure}
\begin{center}
\includegraphics[width=0.6 \columnwidth, clip=true, angle=00]{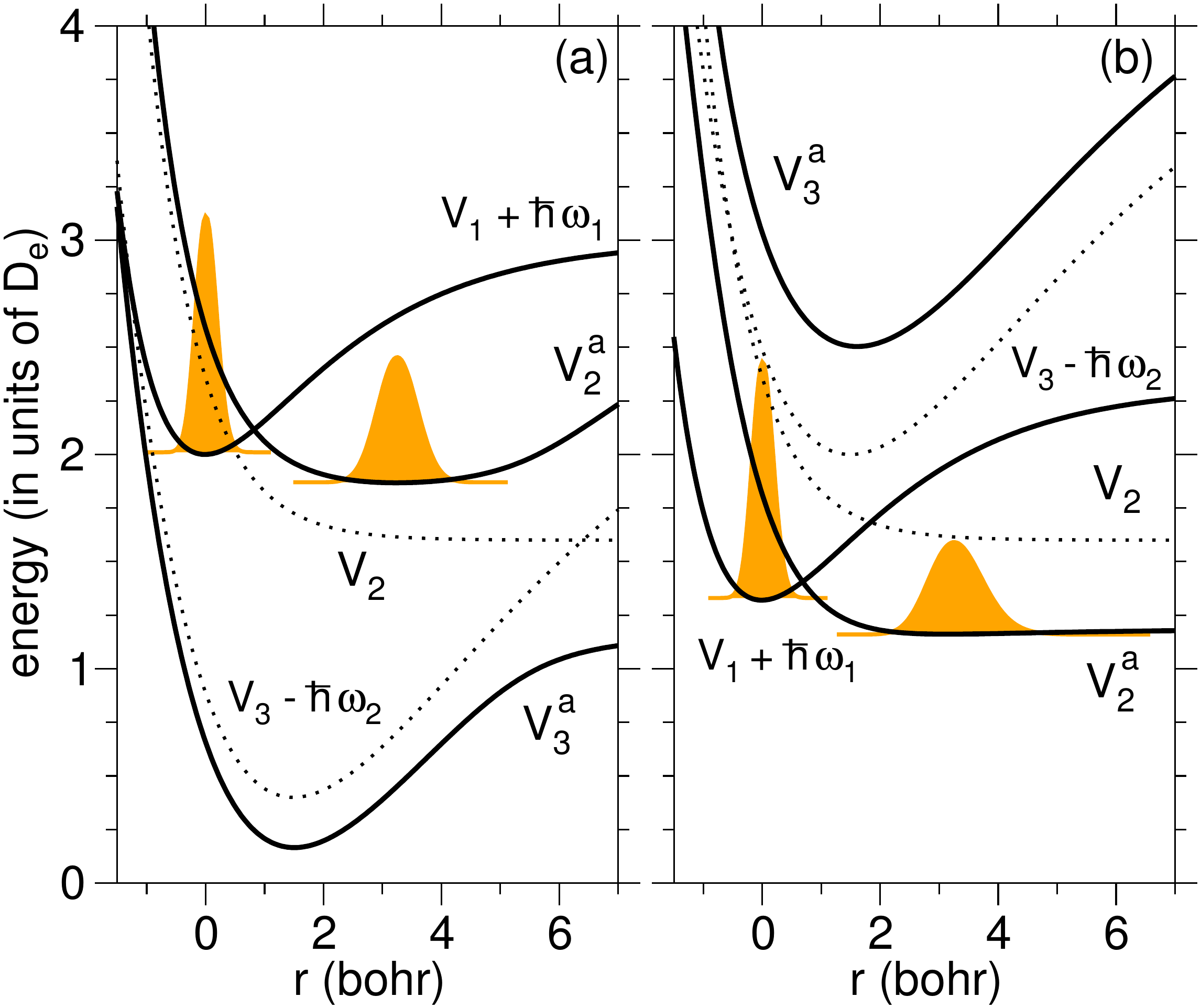}
\includegraphics[width=0.6 \columnwidth, clip=true, angle=00]{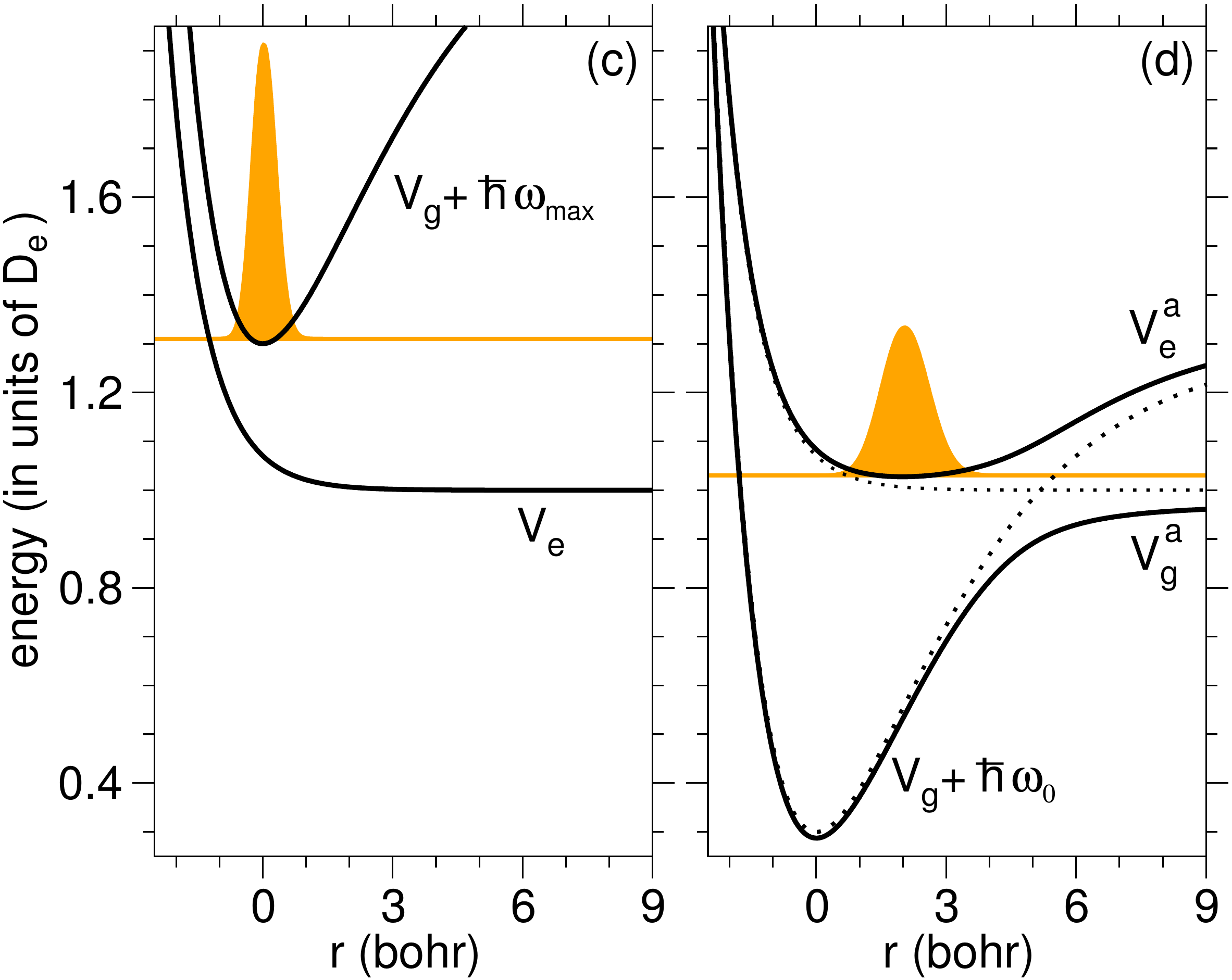}
\caption{Different scenarios of the LAMB scheme using one and two pulses. 
Dotted lines are the molecular potentials and solid lines are the potentials 
in the presence of the field (the LIPs plus the uncoupled $V_1$ potential). 
In (a) and (b) we show LAMB schemes using a chirped and a transform-limited 
pulse, blue-shifted or red-shifted with respect to the $V_2 \rightarrow V_3$ 
transition, respectively.
In (c) and (d) the mechanism of LAMB with a single chirped pulse 
is represented.
 At initial times the pulse frequency must be blue-shifted from the 
photodissociation band, while at later times, the red shifted frequency, 
after sweeping all the photodissociation band, sets the new "equilibrium" 
bond distance. (From J. Chem. Phys. 134, 144303, fig.1)}
\label{LAMBsch}
\end{center}
\end{figure}

More natural LAMB implementations are possible when $V_2$ is coupled
directly to the initial state $V_1$\cite{LAMB2,LAMB3}. 
Fig.\ref{LAMBsch} outlines both two-pulse as well as one-pulse scenarios.
In the first case the LAMB process implies the following mechanism:
Initially $\epsilon_2(t)$ is switched on, with an off-resonant 
frequency that prepares the LIP $V_2^{a}$ with a LIAC between $V_3$ 
and $V_2$ at the desired bond length. 
Then, while $\epsilon_2(t) = \epsilon_0$ remains constant,
another pulse, $\epsilon_1(t)$, moves all the population from
$V_1$ to $V_2$, which in the presence of $\epsilon_0$ is $V_2^a$.
This electronic absorption can proceed rapidly, using an ultrashort
transform-limited pulse that generates a nuclear
wave packet moving in the LIP.
In this case we talk of a vertical wave function transfer (VWT).
Or it can be quasi-static, using a chirped pulse, in which we talk of an
adiabatic transfer.
In both cases full population inversion 
requires pulse bandwidths 
$\Delta\omega_1$, 
at least as large as the absorption band, $\Delta\omega_{FC}$.
In fact, in the quasi-static case, the chirp typically needs to span
an even larger bandwidth. 
Fig.\ref{LAMBsch} (a) and (b) show how the shape of $V_2^a$ 
is influenced
by the choice of $\omega_2$, blue-shifted or red-shifted
with respect to the $V_2 \rightarrow V_3$ transition. In the first
case there is properly a LIAC and the bond length in $V_2^a$ is better defined.
In the second case there is no proper LIAC and the control is mostly
done by Stark effect. Then the LIP is much flatter and it is more difficult
to achieve adiabatic population transfer.

On the other hand, it is possible to use a single chirped pulse, $\epsilon (t)$, 
responsible for both roles: the adiabatic transfer and 
the formation of the LIP. 
The basic mechanism is explained in Fig.\ref{LAMBsch} (c) and (d). 
At initial times
$\omega(0)$ must be blue-shifted from the the absorption spectrum to the 
dissociative state $V_e$. Slowly sweeping through the photodissociation band
the population is transferred in a quasi-static way, with the wave packet
always located at the bottom of the LIP. Then the chirp must
sweep through all the emission spectra. The final value of the frequency,
$\omega_0$, defines the LIAC and the bond length\cite{LAMBc1}. If $\omega_0$ 
is not small enough, $V_e(R_c) - V_g(R_c) - \hbar\omega_0$ will sit at short
$R_c$. On the other hand, when $\hbar\omega_0 \sim D_e$ (where $D_e$ is
the bond energy in $V_g$), the LIAC is at infinite distance.
One can therefore measure the required bandwidth scaled with respect to
$D_e$. Using a single pulse, $\Delta\omega$ must be of the order
of $D_e$.

In comparison with a VWT process a typical LAMB process requires $10$ to $100$
more energy (integrated pulse amplitude or peak amplitude times duration) 
from the pulses. 
The extra energy is mainly used to deform the potential.
This pays off in the fact that the molecular properties
associated to the wave packet dynamics, is entirely governed by the
field parameters. In particular, any {\em trajectory} in the "chirp function"
$\omega(t)$ entails different excursions of the average internuclear
distance, or bond length.
For instance, any time-symmetric function $\omega(t)$ leads to fully 
reversible bond elongations that mimic a single period of a classical 
molecular vibration,
with both the amplitude and frequency of the vibration being externally
controlled\cite{LAMBc3}. If $\omega(t)$ is periodic, the 
inverse of its period will be the "frequency" of this LIP-supported 
vibration\cite{LAMBc2}.

\begin{figure}
\begin{center}
\includegraphics[width=0.7\columnwidth, clip=true, angle=00]{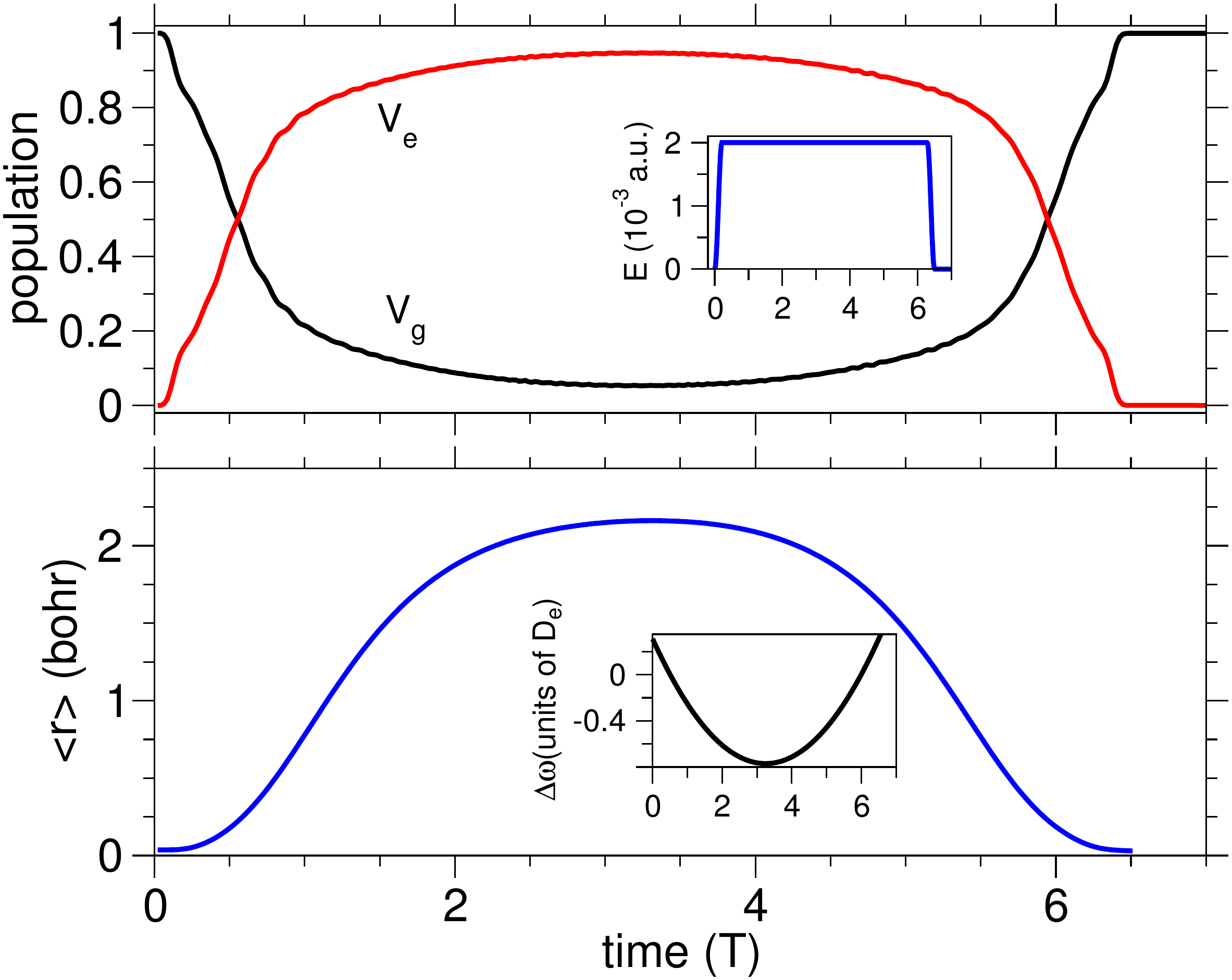}
\includegraphics[width=0.7\columnwidth, clip=true, angle=00]{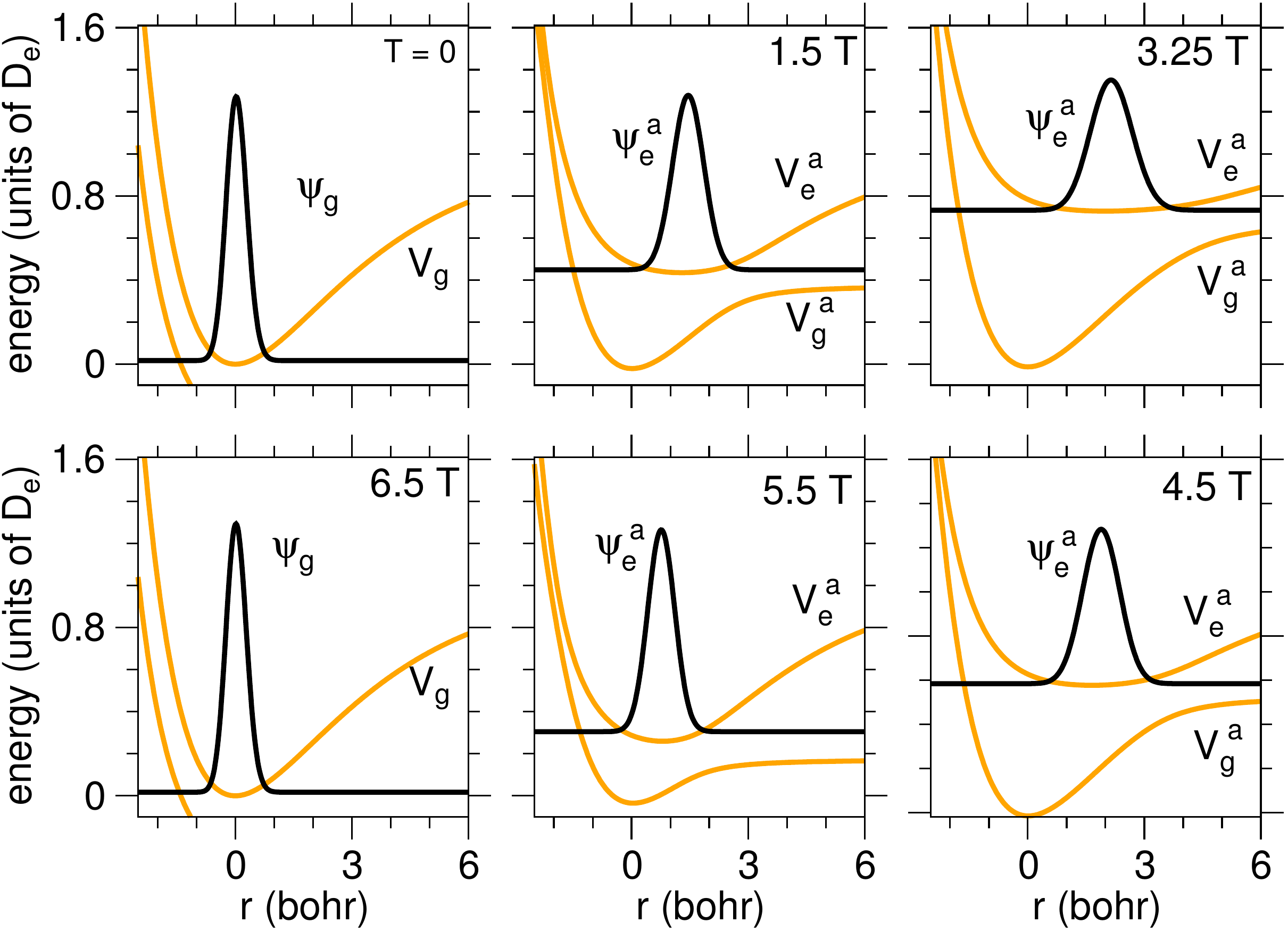}
\caption{Dynamics of the system under a parabolic chirped pulse in adiabatic 
conditions. Time units are scaled with respect to the fundamental vibrational 
period in the ground potential, $T$, whereas the pulse duration is
$\tau = 6.5 T$. (top) The dynamics of electronic populations 
and bond length. Insets show the pulse shape (upper inset) and the laser 
detuning (lower inset). (bottom) Snapshots of the adiabatic wave function, 
remaining in the bottom of the LIP at all times. By chirped induced 
population inversion, $V^{a}_{e}$ correlates at initial and late times 
with the initial potential, $V_g$. 
(From Phys. Rev. A 82, 063414, Fig.2.)}
\label{LAMBsym}
\end{center}
\end{figure}

\begin{figure}
\begin{center}
\includegraphics[width=0.9\columnwidth, clip=true, angle=00]{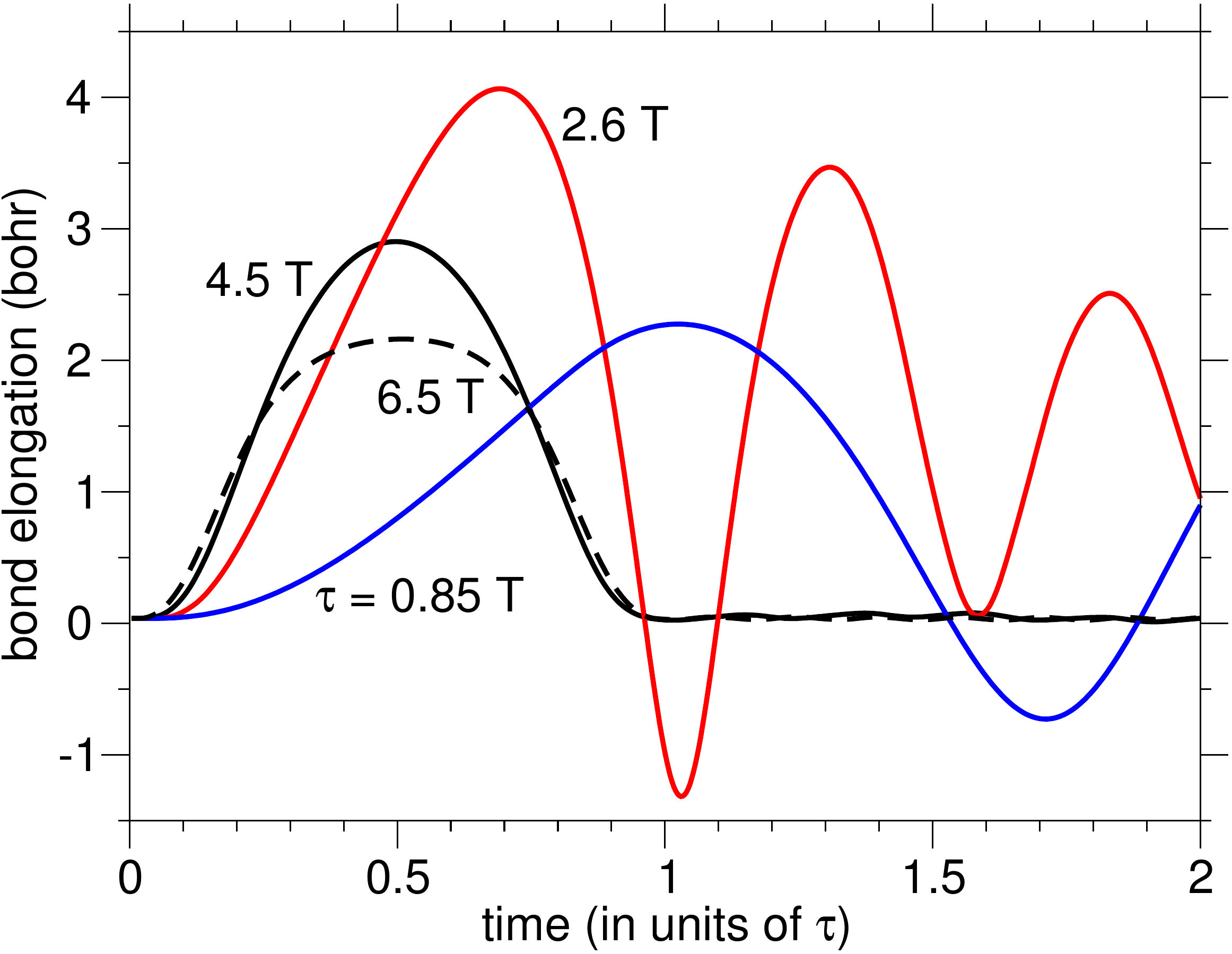}
\caption{Bond dynamics as a function of time scaled with the pulse
duration $\tau$. In the strong adiabatic regime ($\tau \ge 6T$), the 
time-scaled bond dynamics are similar and follow the same curve. 
Before fully adiabatic conditions ($\tau = 4.5T$), the dynamics is still 
time-symmetrical but the bond stretches more at half the pulse duration. 
For shorter pulses the bond dynamics is anharmonic. The wave packet gains 
momentum, and it is no longer attached to the bottom of the LIP. The bond 
can be further stretched, and as the chirp is reversed ($t/\tau > 0.5$), 
the wave function returns to the initial potential as a vibrationally excited 
wave packet. If the pulse is too short, however, the wave packet has not 
time to reach the classical turning point before the chirp reverses and the 
LIP converges to the initial potential. 
(From J. Phys. Chem. A 116, 2691, Fig. 3.)}
\label{LAMBvib}
\end{center}
\end{figure}

In Fig.\ref{LAMBsym} we show a numerical example
of the reversible control of the bond elongation. It can be observed 
that the nuclear wave function is always in the ground state of the LIP.
The exchange of kinetic energy into zero energy of the LIP needs 
time to allow the nuclear wave function to adjust its width to the width
of the potential.
Hence, only "slow" vibrations can be adiabatically controlled,
where the period of the vibration is much larger (typically $5$ or more times
larger) than the characteristic vibrational period of the ground state, $T$.
For a Morse oscillator $V = D_0 (1 - e^{-\beta (R-R_0)} )^2$ of reduced mass 
$m$, $T = \sqrt{ 2m\pi^2 / D_0 \beta^2 }$. 
If $\omega(t)$ changes faster than this characteristic period of time the
process is not fully adiabatic. As Fig.\ref{LAMBvib} shows, the breakdown of 
adiabaticity implies that the nuclear wave function receives some kinetic 
energy, such that the motion of the bond length depends partially on the LIP 
equilibrium geometry determined by $\omega(t)$, and partially on the 
vibrational motion of the nuclear energy on this LIP\cite{LAMBc4}. 
When the chirp is reverted one prepares a highly excited vibrational
wave packet in the ground potential.
However, if $\omega(t)$ varies too fast,
the wave packet has little time to move from its initial position and
the gained vibrational energy is small.

The simplest procedure to prepare large amplitude vibrations (albeit 
with uncontrollable periods) is to use a two-pulse LAMB scheme.
While the control pulse prepares the LIAC and determines the equilibrium
bond distance, one can rely on VWT processes to excite the wave function 
from $V_1$ to the LIP. Some control on the amplitude of the vibration can be 
achieved by selecting the timing and frequency of the ultrashort 
transform-limited pulse\cite{LAMB4}, as shown in the next section.

\section{The Role of the Electronic Charge} 

While the nuclear wave function encodes the molecular structure,
given by the shape or the geometry, the electronic distribution is
responsible for the chemical properties. In particular, the dipole moment
typically provides simplified information regarding the distribution
of charges. Based on the Born-Oppenheimer approximation, in the previous 
section we have focused on the control of the nuclear wave function
by means of LIPs. Here, we analyze the role of the electronic wave function
and of electron-nuclear dynamics\cite{eLIP1,eLIP3,eLIP4}.

In a LAMB process, the total wave function of the system is a coherent 
superposition of both nuclear and electronic wave functions\cite{LAMBc1},
\begin{equation}
\Psi (R,q,t) = \phi_{g}(R,t)\Xi_{g}(q;R)+\phi_{e}(R,t)\Xi_{e}(q;R)
\end{equation}
However, in a fully adiabatic evolution, the nuclear wave packet in 
$V_g$ and $V_e$ have the same shape, 
$\phi_{g} (R,t) \propto \phi_{e}(R,t) \propto \phi^{a}(R,t)$, 
so that we can write
$$\Psi (R,q,t) = \phi^{a}(R,t)\left[a_{g}(t) \Xi_{g}(q;R)+ a_{e}(t)\Xi_{e}(q;R)\right]$$
\begin{equation}
 = \phi^{a}\Xi^{a}(q,t;R)   \label{sepstate}
\end{equation}
where $\Xi^{a}(q,t;R)$ is the dressed electronic wave function that gives
the LIP force field. The total wave function is thus separable and 
not an entangled state of nuclear and electronic states. 
%
Since the total wave function in the
LIP is a single Born-Oppenheimer product, there is perfect correlation
between the electronic and nuclear motion. Notice that the changes of
the electronic wave function are externally controlled: they do not
rely on dynamical phases as in superpositions of different electronic
states. Adiabaticity is required for the single product wave function 
to faithfully represent the dynamics. Hence the changes in the LIP must be 
slower than the typical time-scale of the nuclear dynamics. The
perfect correlation of electronic and nuclear motion in the LIP 
is only possible because the electron dynamics occurs in the time-scale of
the nuclear dynamics.

In order to analyze the type of electron changes conveyed by the
transformation of the LIP and the possible
effects of electron-nuclear motion, one needs to use a model
that allows integrating the full time-dependent Schr\"odinger equation 
beyond the Born-Oppenheimer
approximation. In the following, we use a well-known $1+1$D Hamiltonian
often employed to characterize the dynamics of
the molecular Hydrogen cation under strong fields\cite{eLIP1,eLIP2,eLIP3,eLIP4},
\begin{widetext}
\begin{equation}
{\sf H} =
-\frac{\hbar^2}{2\mu_{e}} \frac{\partial^{2}}{\partial z^{2}} 
- \frac{\hbar^2}{M}\frac{\partial^{2}}{\partial R^{2}}
-\frac{1}{\sqrt{\left(z-\frac{R}{2}\right)^{2} + 1}} 
-\frac{1}{\sqrt{\left(z+\frac{R}{2}\right)^{2}+1}}
+\frac{1}{R}+q_{e}z{\epsilon}(t) 
\label{TDSE}
\end{equation}
\end{widetext}
where $z$ is the electron coordinate, $R$ is the internuclear distance,
$M$ is the mass of the proton, $\mu_{e}=2m_e M/(2M+m_e)$ 
is the reduced mass of the electron, which is approximately the electron
mass $m_e$, and $q_{e}=(2M+2m_e)/(2M+m_e)\approx 1$.
It should be noted that this 
Hamiltonian includes all non-adiabatic couplings
(i.e., the Born-Oppenheimer approximation is not used) although the electron
is forced to move along a line defined by the bond axis through the
approximated soft-core Coulomb potential\cite{SC1,SC2,SC3}.
More elaborate Hamiltonians and calculations exist\cite{Kono1,Kono2}.

\begin{figure}
\hspace*{-0.0cm}
\includegraphics[width=1.0\columnwidth, clip=true, angle=00]{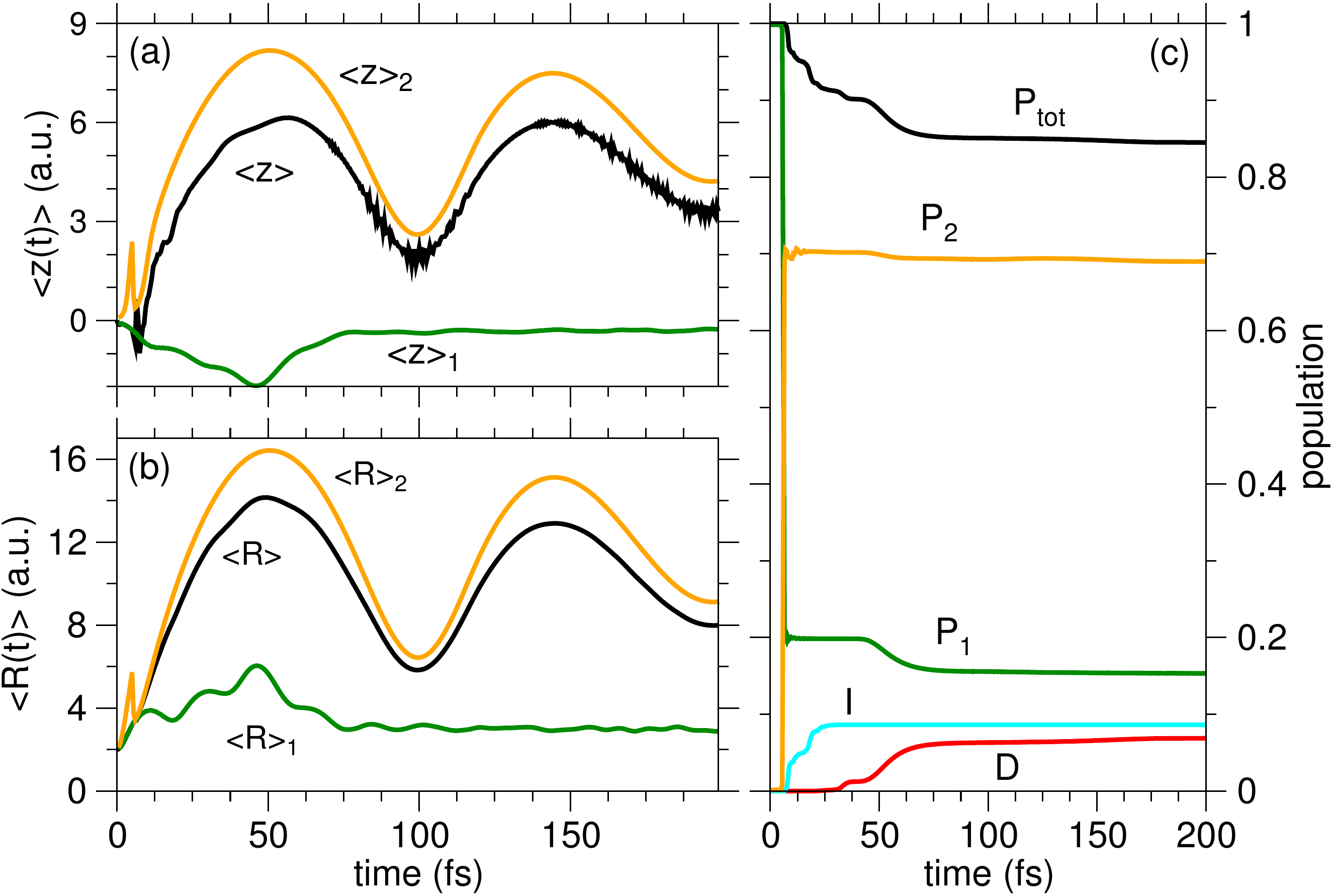}
\caption{(a) Average electron position and
(b) average internuclear distance as functions of time.
(c) Probability as a function of time for dissociation (D), ionization (I)
and the population remaining in the $U_1$ and $U_2$ FIPs, P$_{1}$ and
P$_{2}$, respectively. (J. Chem. Phys. 139, 084306, Fig.2.)}
\label{FIPdyn}
\end{figure}

In Fig.\ref{FIPdyn} we show results of the actual dynamics obtained by 
solving Eq.(\ref{TDSE}) under a strong constant field $\epsilon_0$.
As the initial state, $\Psi(z,R,0) = \varphi(R,0) \psi^{BO}_1(z;R)$, 
we consider a nuclear wave function with the same probability density
as the ground state of H$_2$ in the ground electronic state of the 
ion H$_2^+$, essentially assuming an instantaneous ionization process\cite{eLIP1}.
Since the bond length in H$_2$ is shorter than in H$_2^+$, the ground
state of the former is a nuclear wave packet moving in the ground
electronic state of the cation, $1\sigma_g$. 
An ultrashort pump pulse
of $\tau =1$ fs duration and
carrier frequency $\omega_p = 5.4$ eV, with peak amplitude 
$\epsilon_p = 0.05$ a.u. at $t_0 = 6$ fs is then switched on. 
The duration, frequency, and intensity are chosen to 
maximize population transfer from 
$1\sigma_g$ to the first excited electronic state $1\sigma_u$ 
(henceforth in the section, $V_1$ and $V_2$).
The envelope of the pulse is chosen as a cosine square pulse,
${\epsilon}_p(t) = {\epsilon}_p \cos^2(\pi (t-t_0)/\tau)$ for
$-\tau/2 \le t-t_0 \le \tau/2$.

As in a typical two-pulse LAMB scenario, during all times the control 
field is present. In this calculation we use a {\it DC} electric
field $\epsilon_0 = 0.015$ a.u. Then the dynamics proceeds in the
so-called field-induced potentials (FIPs) $U_1$ and $U_2$, which are
analogous to the LIPs but using constant fields.
$U_1$ shows bond softening. For large fields ($\epsilon_0 \ge 0.03$ a.u.) 
the wave packet kinetic energy can be above the ground state bond energy.
On the other hand $U_2$ shows bond hardening. Indeed, in the presence
of an external field, H$_2^+$ (or any other symmetrical molecular cation)
has charge resonance states\cite{Bandrauk1,Bandrauk2}. 
The transient dipole increases with distance and it is possible to
stabilize the molecule at very large bond lengths.
For the chosen parameters, the population in $V_2$ 
is $\sim0.7$ after the pump pulse, 
while the ionization and dissociation probabilities are
both below $10$\%.
Using more intense DC fields (${\epsilon}_{0} > 0.04$ a.u.),
the ionization and dissociation probabilities increase.

\begin{figure}
\hspace*{-0.0cm}
\includegraphics[width=1.0\columnwidth, clip=true, angle=00]{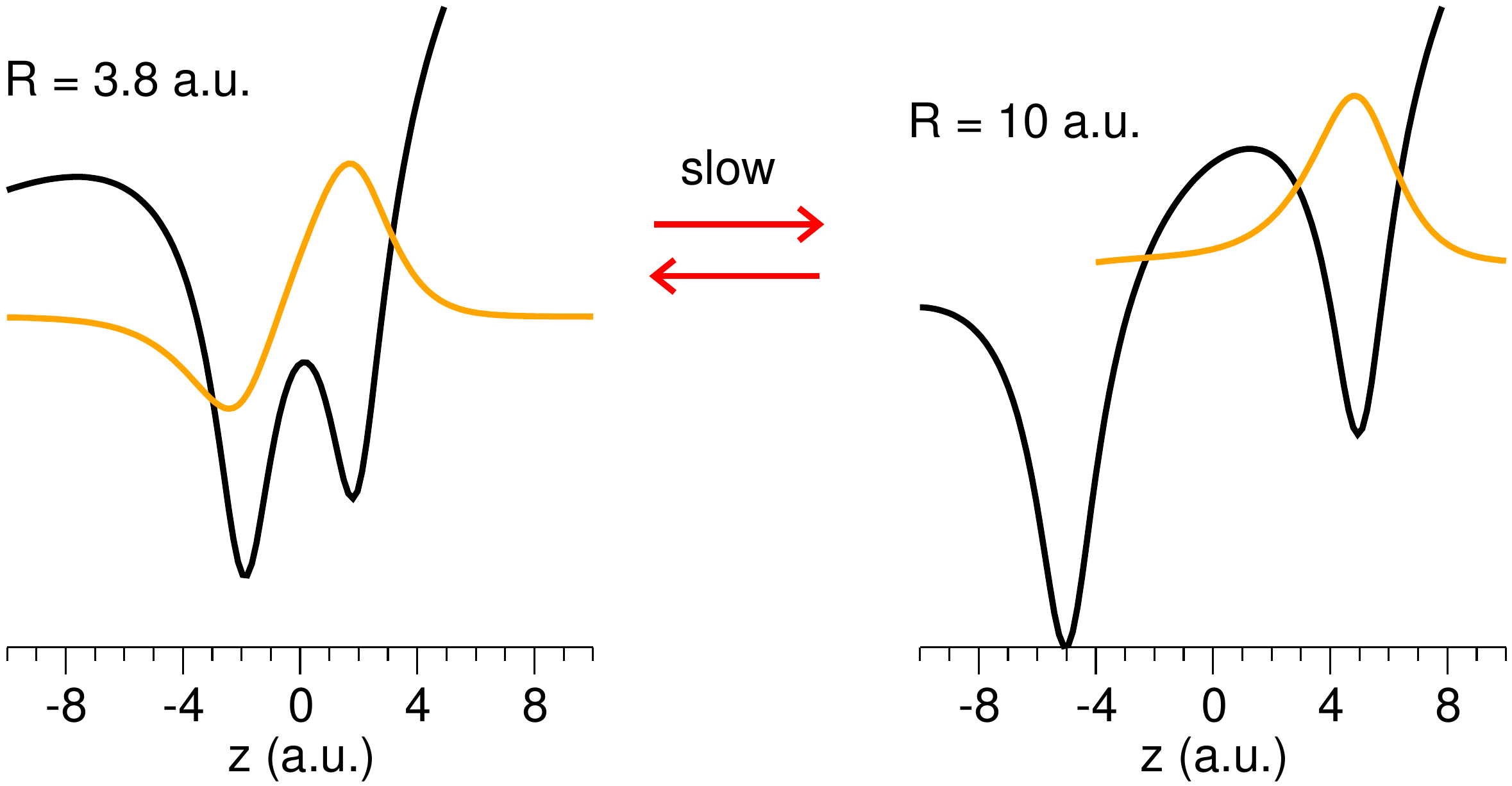}
\caption{Sketch of the success of the mechanism for the creation of the 
dipole starting from the excited FIP. We show slices of the wave function 
and the soft-core Coulomb potential coupled with the field for two different 
internuclear distances which correspond to the initial state and at maximum 
bond stretch. Initially the electron is mainly localized on the right 
atomic well. As the protons separate, as long as the energy of the wave 
function is below the ionization barrier, the electron remains with the 
right proton and moves in the time scale of the vibrational motion.
(From J. Phys. B 48, 043001, Fig.7.)}
\label{FIPmech}
\end{figure}

Fig.\ref{FIPdyn} shows how the motion of the electron and
protons is clearly correlated, with the electron departing from between
the two protons, to the right proton (the one at positive $z$) as the
protons move apart. The period of both motions is practically the same.
The mechanism under this correlation is shown in Fig.\ref{FIPmech},
where we show different snapshots of the soft-core Coulomb potential
sliced at the internuclear distance where the probability of finding
the nuclear wave function is larger. Also shown is the electronic
wave function. As the bond enlarges the electron follows the right
nuclei.
With the chosen field parameters, the average internuclear distance reaches
$14$ a.u. [Fig.\ref{FIPdyn}(b)] while the electron average distance
and hence the electric dipole reaches $6$ a.u. [Fig.\ref{FIPdyn}(a)].
In fact, the wave packet reaches quite longer distances in $z$ and
$R$ than those indicated by the average. Since the FIP is very anharmonic
the dephasing makes the wave packet spread quickly.
The maxima and minima of $\langle R(t)\rangle$ and $\langle z(t)\rangle$ 
become less pronounced after a few periods, 
until the wave packet fully disperses and the average internuclear distance
remains constant. This is the so-called collapsed state. 
For weak DC fields (and weaker bonds), there can be as few as
$2$ periods before the system reaches the collapsed state, 
whereas for stronger fields one can easily observe $10$ periods
of motion. In principle, at larger times one could expect the revival of 
the periodic motion\cite{eLIP3}.
The maximum dipole that can be achieved is also limited by imperfections
in the population transfer between the FIPs,
since the electron
remaining in $U_1$ moves in the opposite direction (with the
left proton) making the average smaller. This effect can be
observed in the averages calculated on each FIP, $\langle z \rangle_j$
and $\langle R \rangle_j$ ($j=1,2$) shown in Fig.\ref{FIPdyn}.

\begin{figure}
\hspace*{-0.0cm}
\includegraphics[width=1.0\columnwidth, clip=true, angle=00]{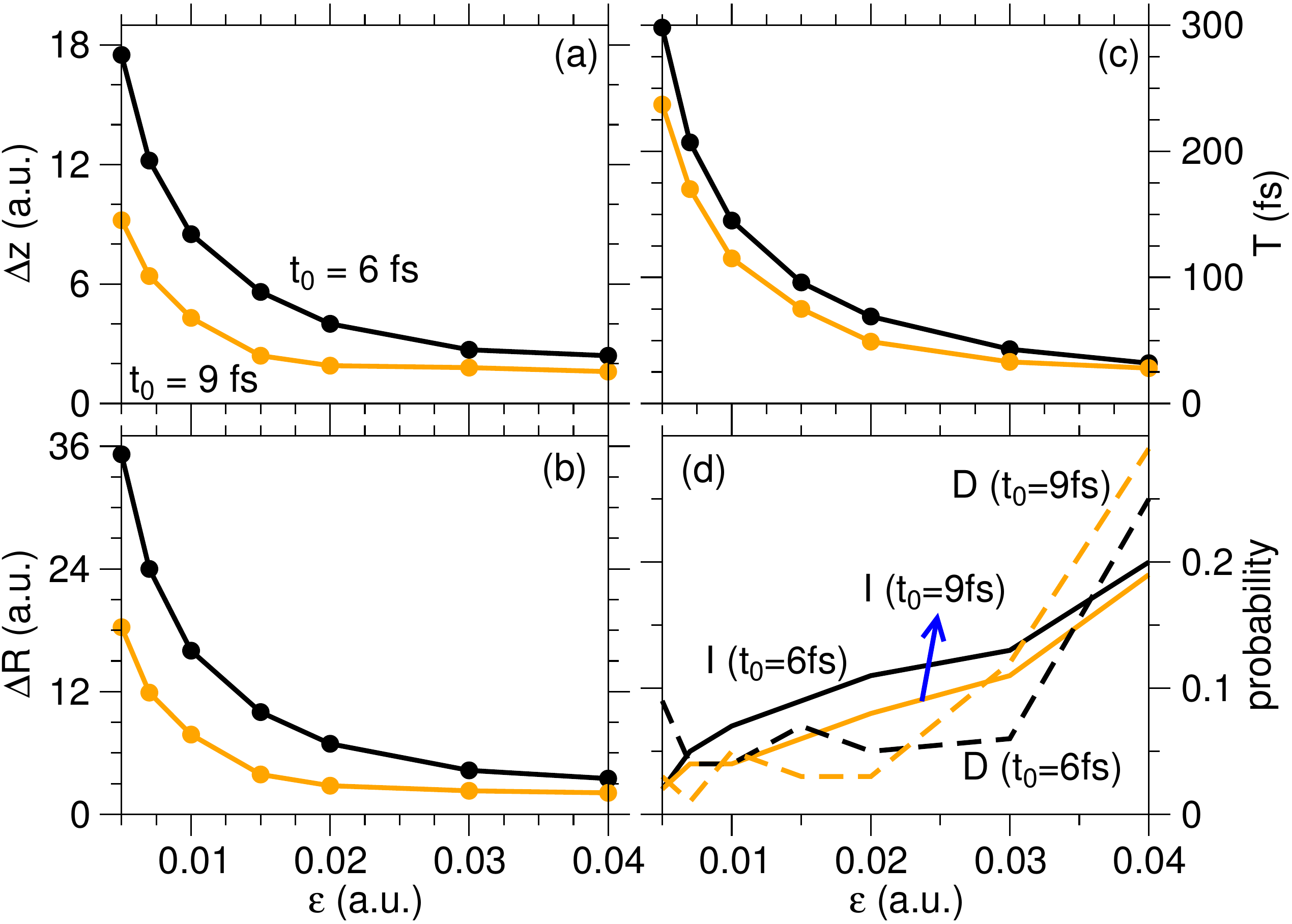}
\caption{(a) Amplitude of the dipole, (b) bond elongation, (c) dipole period,
and (d) probabilities of dissociation and ionization as functions of 
the amplitude of the field $\epsilon_0$ for different time-delays of the 
pump pulse $t_0$. (From J. Chem. Phys. 139, 084306, Fig.3.)}
\label{FIPcontrol}
\end{figure}

Still, extremely large electronic dipoles and bond elongations can be 
achieved.
Defining $\Delta z$ as the range of the dipole motion, $\Delta z \equiv
\langle z(t_{max}) \rangle - \langle z(t_{min}) \rangle$, 
where $t_{max}$ is the time at which 
$\langle z\rangle$ reaches its first maximum and $t_{min}$ is the time at 
which $\langle z\rangle$ is at a minimum during the first period, and
similarly for the bond elongation $\Delta R$, in Fig.\ref{FIPcontrol}
we show how $\Delta z$, $\Delta R$, and the period of the dipole $T$ 
change as functions of $\epsilon_0$ and $t_0$, together
with the probabilities of dissociation 
and ionization. 
For the range of parameters where the ionization and dissociation
probabilities are small, one can generate dipoles that reach as large as $40$
Debye, oscillating at slow vibrational motion. The period of these
dipoles varies from $\sim300$ fs 
down to $25$ fs, i.e., over an order of magnitude.
These correspond to frequencies in the far infrared, from $3$ to
$40$ THz approximately.

It is possible to use a low frequency laser pulse instead of a
constant field, but then 
the pulse must meet very specific conditions\cite{eLIP4}. In particular, its
frequency must be approximately equal to the frequency of the
vibrational motion in the LIP. Otherwise, the correlation between
the motion of the electron driven by the field, and that of the
protons, oscillating in the potential, is not perfect. 
In this scenario, the LAMB dynamics is that of an electron moving
along with one proton as the bond stretches, and hopping to the
other proton as the bond compresses. Basically, we have a proton
loosely attached to an Hydrogen. Depending on the phase of
the electric field, the Hydrogen and proton exchange their role. 
It is crucial that the bond is maximally compressed when the amplitude
of the oscillating electric field is zero as otherwise the electron cannot 
hop from one proton to the other, leading to dissociation.

We have shown that the electron-nuclear correlation is an essential condition 
to create large-amplitude oscillating dipoles.
Although in principle any superposition of electronic states of different
parity, such as $\Psi(z,R,0) = \varphi(R,0)
\left( \psi^{BO}_1(z;R) + \psi^{BO}_2(z;R) \right)/ \sqrt{2}$,
induces an oscillating dipole, the dipole in this case quickly decays
due to the dephasing of the nuclear and electron motion\cite{eLIP3}.
Alternatively, a high frequency laser pulse can be used to drive
the electron in the ground state creating a fast oscillating dipole,
but this dipole can only be small as the excursion length of the
electron in the field (the displacement that the electron can reach before the electric
field changes its sign) is necessary small in the ground potential
under optical driving frequencies. For lower frequencies the excursion
length and the induced dipole could be larger, but tunneling
ionization dominates\cite{eLIP4}.

In summary, we have shown that in order
to create an oscillating electric dipole in a homonuclear diatomic cation
without an oscillating driver one needs (i) to break the symmetry of the
system and (ii) to sustain highly correlated electronic and nuclear motion,
which are guaranteed by the LAMB dynamics.

\section{Control of Photodissociation by Shaping Two Dissociative Potentials}

Strong fields can also be used to control photodissociation reactions in the
adiabatic regime. The first obvious effect is the Stark shift of the
photodissociation bands, thus changing the spectrum\cite{SSS}. 
If several dissociation channels are present,
one can use a pump pulse in combination with a strong nonresonant pulse 
to separate the different channels\cite{diss3}.
However, the field also couples the different excited states so that
the dissociation occurs on a "mixed" channel, that is, on a superposition of 
excited electronic states correlating to different channels\cite{diss3,diss4}.
In this section we review some proposals that we have presented to
achieve control over different reaction observables, such as the
yield and the kinetic energy distribution of the fragments\cite{diss4}, by using
strong nonresonant pulses. In the chosen examples the LIPs are formed between
two dissociative potentials that never cross. Since there are no 
LIACs and no population inversion, the LIPs only show "soft" shaping.

\subsection{Control of photodissociation spectra}

Using an ultrashort pump pulse with carrier frequency $\omega_p$, the 
absorption probability in a photodissociation band quickly decays when the 
absorption is nonresonant. Taking into account the pulse bandwidth 
$\Delta\omega_p$, in the absence of competing dissociating channels,
the photodissociation probability is roughly given by \cite{Meyer}
\begin{equation}
P_j(\omega_p) \sim \exp\left[-\left(\frac{D_{j0} - \hbar\omega_p}{\hbar 
\Delta\omega_p} \right)^2 \right]
\label{dissesp}
\end{equation}
where $D_{j0} = V_j(R) - V_0(R)$ is the energy gap at the Franck-Condon
(FC) region. 
However, using a strong non-resonant field ${\epsilon}_S$ in addition to the 
pump pulse, the electronic states are Stark-shifted. 
Under the same approximations we obtain a similar expression
where instead of $D_{j0}$ one needs to use the Stark-shifted energy gap
\begin{equation}
D^p_{j0}({\epsilon}_S) = V^a_j(R) - V^a_0(R) \approx D_{j0} - \frac{1}{4}
\left( \alpha_{jj} - \alpha_{00} \right) {\epsilon}_S^2
\end{equation}
The position of the photodissociation bands corresponding to different 
electronic channels can therefore be controlled. The control can be more
effective when the polarizabilities $\alpha_{jj}$ have different signs
for different electronic states, such that ${\epsilon}_S$ can both 
blue-shift and red-shift the different bands of the spectra.

\begin{figure}
\centering
  \includegraphics[width=0.85\columnwidth,]{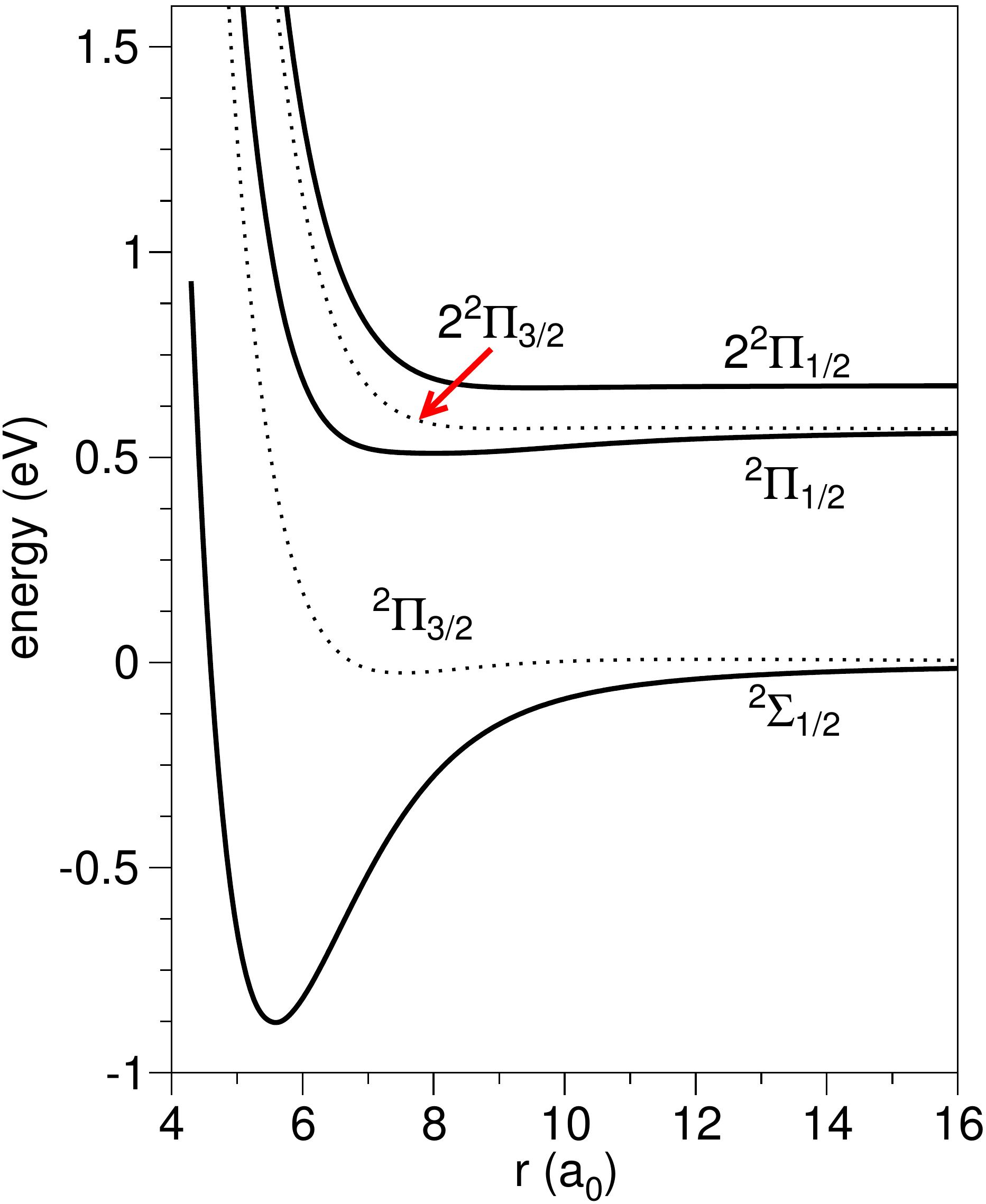}
  \caption{Potential energy curves for ICl$^-$. The ground state is 
$^2\Sigma_{1/2}$. Up to 3.0 eV, the only two excited states that 
are coupled to the ground state by transition dipole moment are
 $^2\Pi_{1/2}$ and 2$^2\Pi_{1/2}$. (Adapted from J. Chem. Phys.
130, 124320, Fig.1.)}
\label{icl}
\end{figure}

\begin{figure}
\centering
  \includegraphics[width=0.85\columnwidth,]{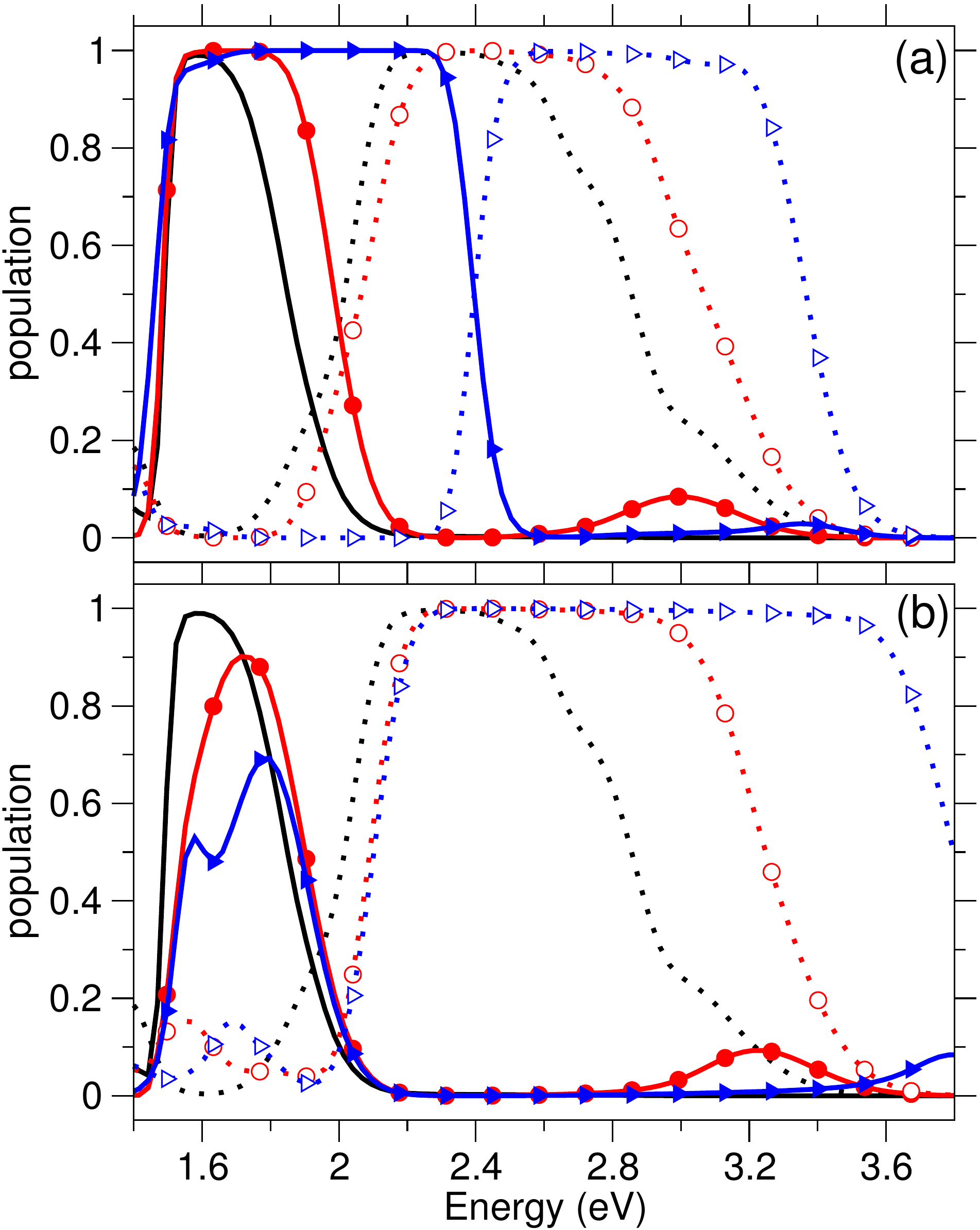}
  \caption{Photodissociation spectra in the presence of the control pulse 
using the counterintuitive sequence with (a) $\epsilon_{S}(t)>0$ and 
(b) $\epsilon_{S}(t)<0$. The keys of the symbols are the following: 
Circles are results with peak amplitudes $\epsilon_{S}= \pm 0.02 e/a^{2}_{0}$, 
triangles are results with Fig $\epsilon_{S} = \pm 0.03 e/a^{2}_{0}$, 
and lines with no symbols are results with $\epsilon_{S}=0$. 
The solid line gives dissociation in $V_{1}$ and the dotted lines 
dissociation in $V_2$. 
The pulse parameters are given in the text. 
(Adapted from J. Chem. Phys. 130, 124320, Fig.6.)}
\label{icl1}
\end{figure}

As a numerical example we consider control on the photodissociation of
the molecular ion ICl$^-$\cite{icl-} shown in Fig.\ref{icl}. 
Assuming the molecule is aligned with ${\epsilon}_S$
(thus the symmetry rules associated to parity can be violated),
the first two excited states, $^2\Pi_{1/2}$ (henceforth $V_1$) and 
2$^2\Pi_{1/2}$ (henceforth $V_2$), are
dipole allowed from the ground state $^2\Sigma_{1/2}$ ($V_0$) and are the only
accessible states below $3$eV or $400$nm. 
The two excited states, on the other hand, are too far apart to
be reached by a single pulse (unless it is an attosecond pulse) and 
do not cross at different internuclear distances, so that internal 
conversion is negligible.

In Fig.\ref{icl1} we show the photodissociation spectra when the pump
pulse is a $400$ fs pulse of a TW/cm$^2$ in the range where the two bands
are observed. The spectra barely depends on the parameters of the pump pulse.
However, it changes considerably when a very strong nonresonant ${\epsilon}_S$
pulse is included. 
For the results of Fig.\ref{icl1} we have assumed that its frequency is 
negligible, that is, we have employed half-cycle pulses.
The full width at half maximum (FWHM) of ${\epsilon}_S$ was chosen as $400$ fs, 
which implies that the control pulse lasts about twice the duration of the 
pump pulse. 

Interestingly, although for the chosen parameters
$\alpha_{22}$ and $\alpha_{33}$ are both positive and of similar value,
the spectral bands after ${\epsilon}_S$ are not only blue shifted. Indeed
the high frequency edge of the band is blue shifted so that, for instance,
with ${\epsilon}_S = 0.03$ a.u. the band for $V_1$ overlaps the spectral
window of the band for $V_2$ in the absence of the pulse.
However, the low-frequency edge of the bands remain practically the same
as without the Stark-shift. This is because the pump pulse is also acting
when ${\epsilon}_S$ is small (either at the head or trail of the pulse, depending on
the time delay between the pulses), so that in the spectra one essentially
records the yield of photodissociation for all possible Stark-shifts,
obtained with ${\epsilon}_S(t)$ ranging from zero to its peak amplitude.
A simple Stark-shift of the whole band would be observed for much larger
control pulses when ${\epsilon}_S(t)$ straddles ${\epsilon}_p(t)$, such that
${\epsilon}_p(t)$ is switched on after and switched off before the control pulse.

\subsection{Control of photodissociation yields.~~}

\begin{figure}
\centering
  \includegraphics[width=0.80\columnwidth,]{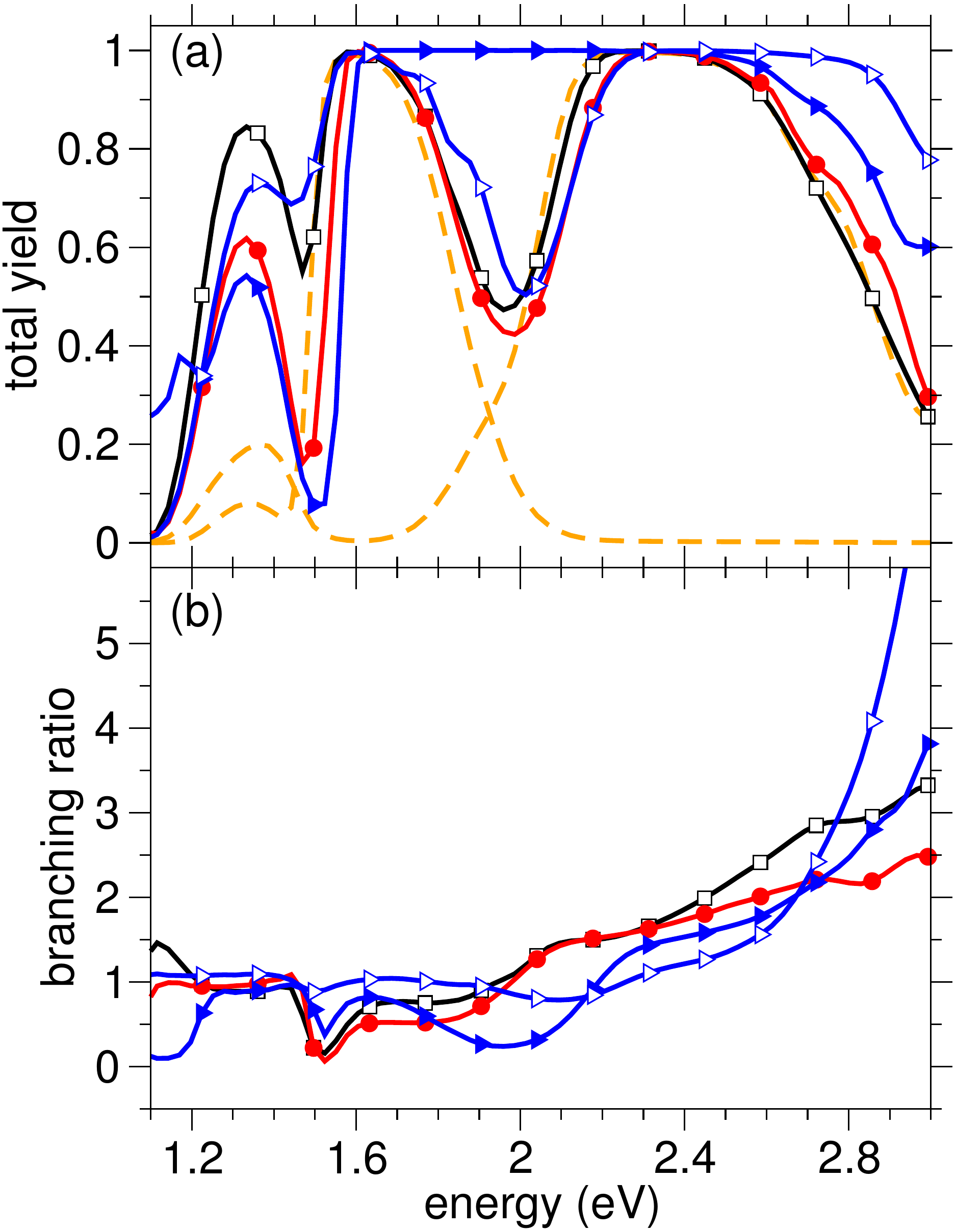}
  \caption{(a) Photodissociation spectra in the presence of the control 
pulse using the intuitive sequence for different control peak amplitudes 
and (b) the resulting branching ratio $\chi = P_{2}/P_{1}$. 
In (a) we give the total yield of dissociation ($P_{1}+P_{2}$). 
The peak amplitude of the control field for the different results 
is the following: In dashed line $\epsilon_{S}=0$; in black line 
with squares $\epsilon_{S}=−0.01$; in red line with circles 
$\epsilon_{S}=0.02$; in blue line with solid triangles $\epsilon_{S}=0.03$ 
and in blue line with empty triangles $\epsilon_{S} =−0.03e/a^{2}_{0}$. 
All other parameters in the simulations are given in the text.
(Adapted from J. Chem. Phys. 130, 124320, Fig.7.)}
\label{icl2}
\end{figure}

In Fig.\ref{icl1} we showed the photodissociation spectra, with the 
expected band shifts.
However, as indicated in Sec.2, when a strong pulse couples
two states, the nature of the states change by virtue of the polarizability.
In particular, there will be some probability that a nuclear wave packet
dissociating at the band corresponding to $V_1$ will actually be in the
excited $V_2$ state and viceverse. For an approximately constant 
$\Omega_S = \mu_{12}(R) {\epsilon}_S(t)$ the probabilities are roughly 
given by 
\begin{equation}
\chi = \frac{P_2}{P_1} \sim \left( \frac{\Omega}{2\Delta} \right)^2.
\label{ratio}
\end{equation}
where $\Delta$ is the energy gap between the electronic states.
The first question, then, is why in Fig.\ref{icl1} we observe only one
product, the electronic channel corresponding first to $V_1$ and then to
$V_2$, at the two photodissociation bands. The reason is because 
we used a constant $\mu_{23}(R)$ and the control pulse was switched off 
slowly. Then the effect of the polarizability that occurred as 
${\epsilon}_S(t)$ raised was reverted when the pulse decayed. 
Eq.(\ref{ratio}) was only operative when the control pulse was switched on
leading to transient effects, but it does not affect the asymptotic results.

However, when either $\mu_{12}(R)$ or ${\epsilon}_S(t)$ decay abruptly
(such that $\Omega_S(t)$ decays abruptly) then the effect of the polarizability
is no longer time-symmetrical and there are interesting effects that 
depend on the dynamics, that is, on the choice of the pulse sequence.
We considered two cases: the PS sequence where ${\epsilon}_p(t)$ overlaps
the head of ${\epsilon}_S(t)$ and the SP sequence, where ${\epsilon}_p(t)$ overlaps
the trail of ${\epsilon}_S(t)$. In the results shown in Fig.\ref{icl2} we
assumed that $\mu_{12}(R)$ decayed abruptly. Similar results would be
obtained using ${\epsilon}_S(t)$ with fast decay.

In the SP sequence the Franck-Condon excitation proceeds between
$V_0$ and the spectrally chosen (by $\omega_p$) excited molecular state $V_e$
($e = 1,2$) but the dissociation occurs in the asymptotic region of the 
molecular potential, leading to selective dissociation. If the chosen potential 
$V^a_e$ is {\it e.g.} $V^a_2$, then one collects all the fragments in the 
molecular state that correlates with that potential, that is, $V_2$,
\begin{equation}
\psi_0(R,t) \stackrel{S}{\longrightarrow} \psi^a_2(R,t) 
\stackrel{P}{\longrightarrow} \psi_2(R,t)
\end{equation}
where $\psi^a_j$ is the wave function initially in state $j$ but of mixed 
electronic character by virtue of the Stark pulse.
The SP sequence leads to the results shown in Fig.\ref{icl1}.
Conversely, in the PS sequence the Franck-Condon excitation
occurs mainly in the excited molecular state $e$, selected by $\omega_p$,
but the dissociation occurs in the asymptotic region of $V^a_j$, 
leading to mixed dissociation.
For instance, if we initially excite $V_1$,
\begin{equation}
\psi_0(R,t) \stackrel{P}{\longrightarrow} \psi_1(R,t) 
\stackrel{S}{\longrightarrow} \psi^a_1(R,t) \propto \sqrt{\chi({\epsilon}_S)}
\psi_1(R,t) + \psi_2(R,t)
\end{equation}
where $\chi$ is roughly given by Eq.(\ref{ratio}).
The labels $2$ and $1$ should be exchanged if we initially excite the 
system in the second photodissociation band, $V_2$.
In Fig.\ref{icl2} we show results of the photodissociation spectra
using this sequence. In this case, the dynamics in the LIP mixes
both dissociation channels. Fig.\ref{icl2}(a) shows the overall yield
of dissociation while Fig.\ref{icl2}(b) shows the branching ratio. Quite
naturally, the band at lower frequency corresponds to excitation 
in the $V_{1}^{a}$ and leads to maximal dissociation in $V_1$,
while the band at higher frequency corresponds to excitation in $V_2^a$
and leads to maximal dissociation in $V_2$.

Clearly, regardless of the pump pulse, the timing of the control pulse
with respect to the pump pulse affects the yield of the
photodissociation reaction. 
However, one needs to work on resonance to create LIACs and fully invert
the electronic populations to have more ample control over the branching
ratios.

\subsection{Control of the kinetic energy distribution.~~}

\begin{figure}
\centering
  \includegraphics[width=1.0\columnwidth]{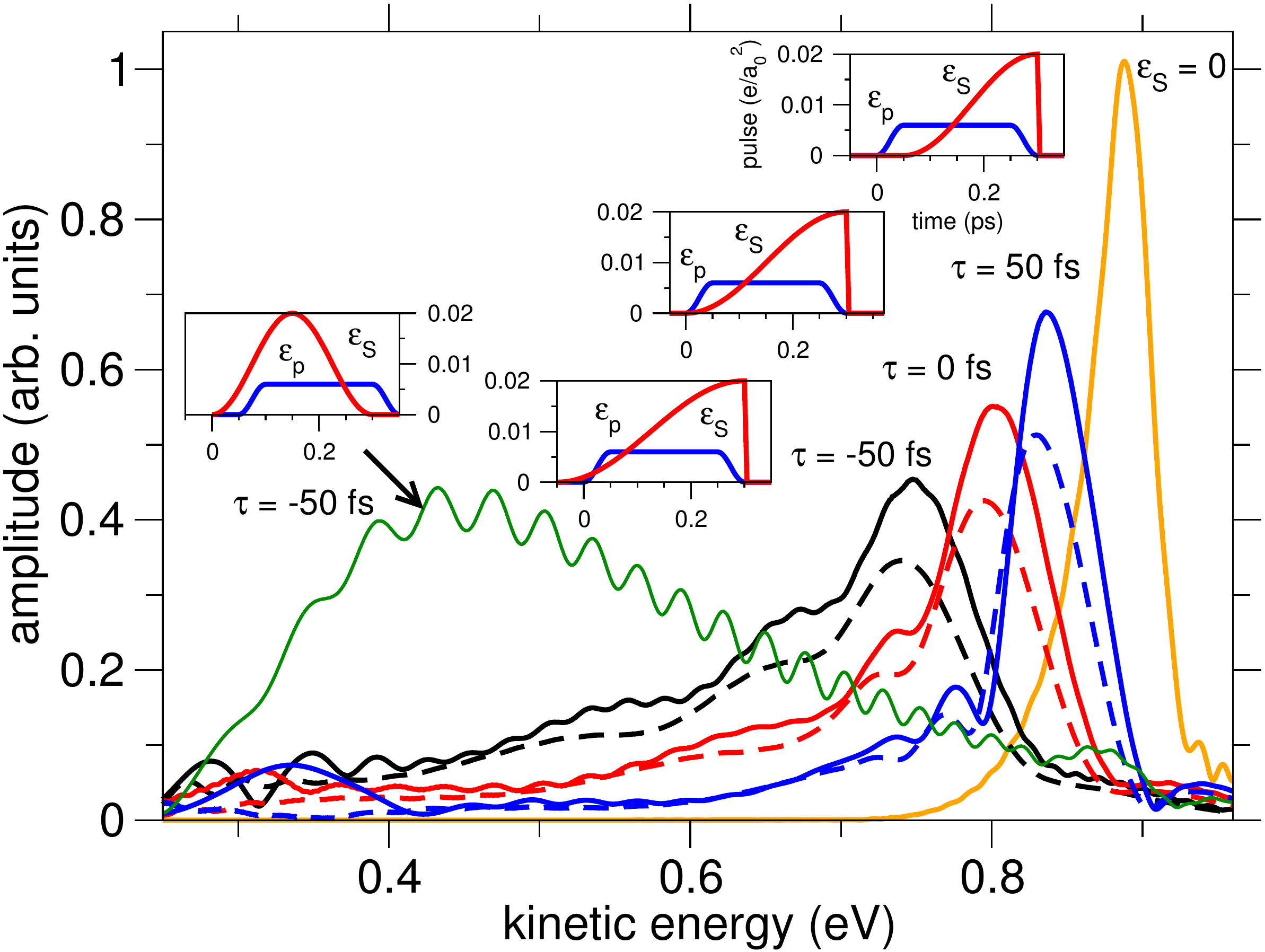}
  \caption{Kinetic energy distributions of the relative motion of the fragments
as a function of the time-delay between the pump and control pulses. The
pulse sequences are shown in insets. The control pulse provokes broadening
in the KED with different asymmetries for the PS and the SP sequences.
In the PS sequence the solid-line curve shows the KED in the electronic 
state $V_2$, and a dashed-line curve shows the KED in the electronic state 
$V_1$. The distributions for the different electronic channels almost overlap 
owing to the dynamics in the same Stark-shifted potential. Because the control
pulse is very intense the probabilities on both channels 
are not very different.
 The peaks of the distributions are always red-shifted with respect to the 
control free case, owing the the blue-Stark shift of the $V_2$ excited 
potential. (From J. Chem. Phys.131, 204314, Fig.4.)}
\label{icl3}
\end{figure}

The use of a strong nonresonant control field can lead to an important
shift of the kinetic energy distribution (KED) of the fragments.
At final time the wave packets
move in either $V_1$ or $V_2$ and the maximum relative speed will
be given by the energy difference between the FC window of
the chosen excited LIP, $V^a_e(R)$, and the asymptotic value of the
molecular potential to which it correlates, $V_e(\infty)$.
If ${\epsilon}_S$ is large, then $V_e^a$ is largely blue shifted. As the
carrier frequency of the pump $\omega_p$, is constant,
the energy difference between the Franck-Condon region and the asymptotic
threshold, which is fixed as $V_e(\infty)$, is smaller.
Therefore, for constant ${\epsilon}_S$ one can achieve red-shifting of the KED.

In addition, using control pulses ${\epsilon}_S(t)$, the Stark-shifted KED
will be very broad, like the photodissociation bands shown in Fig.\ref{icl3}.
Since during the absorption ${\epsilon}_S(t)$ is varying with time,
the KED comprises energies available when ${\epsilon}_S(t) = 0$
(that is, in the absence of control pulse) to energies available when
${\epsilon}_S(t)$ is at its peak value, which are greatly red shifted.

A most interesting effect is the ability to detect a similar KED in both 
excited electronic states in the same experiment. The previous protocol 
of a PS sequence that allows to control the yield of the direct 
photochemical reaction enables this possibility. 
In the results shown in this section we
will use half cycle control pulses ${\epsilon}_S(t)$ ($\omega_S = 0$)
with very fast ($\sim5$ fs) switch off times, implying the sudden change 
of the wave packet from $\psi^a_e(t)$ to $\psi_1(t)$ and $\psi_2(t)$,
both initially opened, and the initially closed electronic channels.
Instead of changing the peak amplitude of ${\epsilon}_S$, the control will
be exerted by changing the time-delay between the pulses.
Using the same ICl$^{-}$ model introduced before, in Fig.\ref{icl3} we show
the control exerted on both KEDs, which are practically identical for both
products.

All the KEDs using the PS sequence exhibit a kind of asymmetry to the red, 
with the low-energy tail much larger than the high-energy tail.
This is because in the PS sequence the absorption induced by ${\epsilon}_p(t)$
occurs mainly when ${\epsilon}_S(t)$ is small,
and thus the potentials are only slightly blue-shifted. Blue-shifted
asymmetries occur using the SP sequence. Then the bulk of the wave
packet is excited when ${\epsilon}_S(t)$ is large, so that the peak of the
KED is now much more red shifted. As explained in the previous section,
in this case the photodissociation is selective and only one final state
is observed.
In the example shown in Fig.\ref{icl3} the peak of the KED is now 
close to $0.4$ eV
implying a larger deviation ($>30$\%) of the relative speed of the fragments.
We observe additional features, like the characteristic structure of
interference due to excitation of the wave packet to the dissociative
potential at different times~\cite{Leinetal}.

In very nonresonant conditions, the manipulation of the KED
relies solely on the ability to Stark-shift the potential at
the Franck-Condon region. In most cases, under reasonable pulse intensities,
the expected effect will be small, as the Franck-Condon window typically
lies in the fast-exponential repulsive wall of the potential and cannot be
largely shifted in the same way that the absorption spectra cannot be
greatly altered.
However, using resonant control pulses between $V_1$ and $V_2$ leads to
a full reshaping of the LIP.
The key then for being able to change the asymptotic KEDs relies on
non-adiabatically disrupting $\Omega_S(t)$ during the wave packet evolution
through the reaction coordinate. Then the energy differences between 
$V_\mathrm{e}^\mathrm{a}(R)$ and the molecular potentials $V_1(R)$ and $V_2(R)$
at the time of the sudden switch off of $\epsilon_S(t)$ will be imprinted in the final
KEDs~\cite{SFC3}.

\section{Conclusions}

In this review we have provided an overview of the theoretical framework
that allows to describe and interpret many dynamical processes of 
diatomic molecules under the influence of strong laser fields, acting
below the threshold of ionization. We have analyzed several theoretical 
proposals using laser schemes that allow to enlarge the bond length of 
molecules, to fabricate slow coherent vibrations, to generate huge
transient dipole moments and to control the yield, branching ratio
and kinetic energy distribution of the fragments after a direct
photodissociation reaction. We have illustrated these proposals with
results taken from some of our contributions in the field.

The experimental realization of some of these proposals is still lacking.
It is typically difficult to find systems where the effect of the
field is strong enough to deform the molecular potentials, yet not
as strong as to completely ionize the molecule.
Some of the proposed schemes also rely on frequency modulation over a
bandwidth larger than what is possible with currently technology. 
More tests are needed, 
particularly in polyatomic molecules, to assess the validity of the 
approximations involved and to discover new ways to minimize the impact of 
ionization.
However, recent experiments have shown that it is possible to achieve
unprecedented control over photodissociation reactions\cite{SFC1,SFC2,SFC3}.
The LIPs provide a new playground where essentially new electronic
states are created and these new chemical species are an open door
to control the chemistry of simple molecules or to increase our
understanding of the molecular dynamics in excited states.

\section*{Acknowledgments}

This work was supported by the NRF Grant funded by the Korean government
(2007-0056343), the International cooperation program (NRF-2013K2A1A2054518),
the Basic Science Research program (NRF-2013R1A1A2061898),
the EDISON project (2012M3C1A6035358),
and the MICINN project CTQ2012-36184.

\end{document}